\documentclass[aps,prd,twocolumn,floatfix,noshowpacs,tightenlines,noshowkeys,superscriptaddress,amsmath,amssymb,
nofootinbib]{revtex4-1}
\usepackage{amssymb,amsbsy,epsfig,color,graphicx}
\usepackage{color}
\usepackage{[longtable}
\usepackage{array}
\usepackage{dcolumn}   % needed for some tables
\usepackage{cellspace}
\usepackage{mathtools}
\usepackage{amstext}
\usepackage{amssymb}
\usepackage{stmaryrd}
\usepackage{stackrel}
\usepackage{graphicx}
\usepackage{esint}
\usepackage[utf8]{inputenc}
\usepackage{float}
\restylefloat{table}
\usepackage{booktabs}
\usepackage{enumitem}
\usepackage{slashed}
\usepackage{hyperref}
\usepackage{etoolbox} % for \appto
\usepackage{lipsum} % for mock text
\usepackage[capitalize]{cleveref}
\usepackage{multirow}
\usepackage[caption=false]{subfig}
\allowdisplaybreaks
\usepackage{soul}

\def\be{\begin{equation}}
\def\ee{\end{equation}}
\def\bea{\begin{eqnarray}}
\def\eea{\end{eqnarray}}

\newcommand{\ba}{\begin{eqnarray}}
\newcommand{\ea}{\end{eqnarray}}

\renewcommand\[{\begin{equation}}
\renewcommand\]{\end{equation}}
\makeatletter

\appto{\appendix}{%
\@ifstar{\def\theequation@prefix{A.}}%
{}%
}
\makeatother

\hypersetup{pdfstartview=FitV,colorlinks=true,linkcolor=midblue,citecolor=midblue,filecolor=midblue,urlcolor=midblue}

\definecolor{midblue}{rgb}{0,0,0.5}
\definecolor{cadmiumorange}{rgb}{0.93, 0.53, 0.18}

\begin{document}

\title{How does a dark compact object ringdown?}

\author{Elisa Maggio}
\email{elisa.maggio@uniroma1.it }
\affiliation{Dipartimento di Fisica, “Sapienza” Università di Roma \& Sezione INFN Roma1, Roma 00185, Italy}

\author{Luca Buoninfante}
\email{buoninfante.l.aa@m.titech.ac.jp}
\affiliation{Department of Physics, Tokyo Institute of Technology, Tokyo 152-8551, Japan}

\author{Anupam Mazumdar}
\email{anupam.mazumdar@rug.nl}
\affiliation{Van Swinderen Institute, University of Groningen, 9747 AG, Groningen, The Netherlands}

\author{Paolo Pani}
\email{paolo.pani@uniroma1.it}
\affiliation{Dipartimento di Fisica, “Sapienza” Università di Roma \& Sezione INFN Roma1, Roma 00185, Italy}

\begin{abstract}
A generic feature of nearly out-of-equilibrium dissipative systems is that they resonate through a set of quasinormal 
modes. Black holes --~the absorbing objects \emph{par excellence}~-- are no exception. When formed in a merger, black 
holes vibrate in a process called ``ringdown'', which leaves the gravitational-wave footprint of the event horizon.
In some models of quantum gravity which attempt to solve the information-loss paradox and the singularities of 
General Relativity, black holes are replaced by regular, horizonless objects with a tiny effective reflectivity. 
Motivated by these scenarios, here we develop a generic framework to the study of the ringdown of a compact object with 
various shades of darkness. By extending the black-hole membrane paradigm, we map the interior of any compact object in 
terms of the bulk and shear viscosities of a fictitious fluid located at the surface, with the black-hole 
limit being a single point in a three-dimensional parameter space.
We unveil some remarkable features of the ringdown and some universal properties of the light ring in this framework.  
We also identify the region of the parameter space which can be probed by current and future gravitational-wave 
detectors. A general feature is the appearance of mode doublets which are degenerate only in the black-hole limit.
We argue that the merger event GW150914 already imposes a strong lower bound on the compactness of the merger remnant 
of approximately $99\%$ of the black-hole compactness. This places model-independent constraints on black-hole 
alternatives such as diffuse ``fuzzballs'' and nonlocal stars.
\end{abstract}

\maketitle

%%%%%%%%%%%%%%%%%%%%%%%%%%%%%%%%%%%%%%%%%%%%%%%%%%%%%%%%%%%%%%%%%%%%%%%%

\section{Introduction}

At the beginning of the last century, the quantum description of atomic interactions had been dramatically 
revolutionized by \emph{precision} atomic spectroscopy. The observation of a small effect such as the Lamb shift 
in the energy levels of the hydrogen atom~\cite{LambShift} had a paramount impact in shaping the development of quantum 
electrodynamics.
Following the historical detection of the gravitational waves~(GWs) from binary 
mergers~\cite{TheLIGOScientific:2016src,LIGOScientific:2019fpa}, black-hole~(BH) 
spectroscopy~\cite{Vishveshwara:1970cc,ChandraBook,1980ApJ...239..292D,Dreyer:2003bv,Berti:2005ys,Kokkotas:1999bd,
Berti:2009kk}
--~i.e., the measurement of the GW spectrum of a BH through post-merger ``ringdown'' 
observations~\cite{Abbott:2016blz,TheLIGOScientific:2016src}~-- may play a similar 
major role in probing the gravitational interaction and fundamental 
physics~\cite{Berti:2015itd,Barack:2018yly,Berti:2018vdi,Sathyaprakash:2019yqt}. 

Indeed, BHs can be considered as the ``hydrogen atom'' of gravity, and their GW spectrum is 
their characteristic footprint.
At variance with the hydrogen atom, though, BHs are \emph{dissipative} systems and their spectrum is described by 
quasinormal modes~(QNMs), each one defined by an oscillation frequency $\omega$ and a damping time $\tau$ related to 
the dissipation due to the GW emission at infinity and the absorption at the horizon.

The defining feature of a BH is its event horizon, i.e., a hypersurface delimiting the spacetime region from which 
no signal can escape. Accordingly, BHs are perfect absorbers at the classical level\pagebreak\footnote{For 
astrophysical BHs, the semi-classical phenomenon of Hawking radiation~\cite{Hawking:1974sw} is negligible.}.
The presence of a horizon provides the key to reading BH physics~\cite{Cardoso:2019rvt}: the strong 
redshift near these objects, the existence of a photon-sphere~\cite{Cardoso:2014sna,Cunha:2020azh}, the number of 
independent ``charges'', their multipole moments, their vanishing tidal deformability, and last but not least their QNM 
spectrum. 

Nonetheless, the presence of an event horizon poses some theoretical and fundamental problems~\cite{Cardoso:2019rvt}, 
the most notable ones being the presence of a curvature singularity inside BHs (where Einstein's theory breaks down) 
and the Hawking information loss paradox (according to which unitarity of quantum mechanics is lost if a spacetime has 
an event horizon~\cite{Hawking:1976ra}). Different attempts to solve 
these long-standing problems have been proposed. Many of them share the common feature that classical BHs are replaced 
by horizonless, singularity-free solutions. This is the case of the ``fuzzball'' 
proposal~\cite{Lunin:2001jy,Lunin:2002qf,Mathur:2005zp}, according to which a large number of regular, horizonless 
microstate geometries with the same asymptotic charges of a BH emerge as solutions to supergravity and string 
theories~\cite{Myers:1997qi,Mathur:2005zp,Bena:2007kg,Balasubramanian:2008da,Bena:2013dka}.
Another example are nonlocal stars~\cite{Buoninfante:2019swn} that emerge in theories with infinite-derivatives in 
which the nonlocality of the gravitational interaction can smear out the curvature singularity and avoid the presence of 
a horizon~\cite{Nicolini:2005vd,Koshelev:2017bxd,Buoninfante:2018rlq,Buoninfante:2018xif,Biswas:2011ar,Frolov:2015bta}. 

In this context, the devising of observational tests to distinguish a classical BH from another dark compact object 
is of utmost importance, and current GW observations (such as GW190814~\cite{Abbott:2020khf}) do not exclude the 
existence of exotic compact objects other than BHs and neutron stars. Arguably, GW spectroscopy is the most direct 
way to test the nature of a remnant formed in the highly-dynamical aftermath of a gravitational merger. Even small 
deviations in the QNMs predicted for a BH in General Relativity might provide indication for new physics, similarly to 
the Lamb shift in the spectrum of the hydrogen atom.

\begin{figure}[t]
\centering
\includegraphics[width=0.45\textwidth]{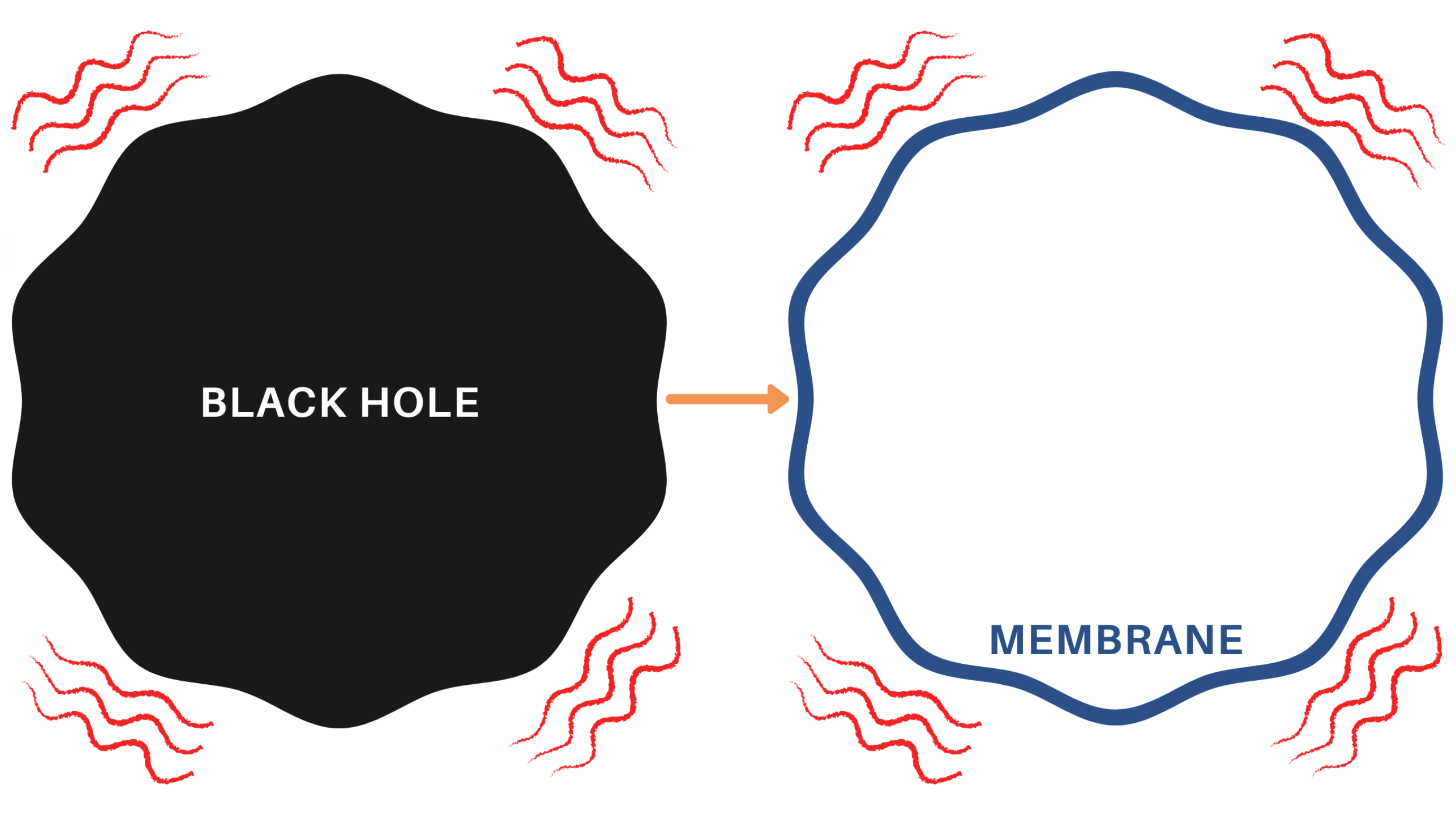}
\caption{Schematic representation of the BH membrane paradigm~\cite{Damour:1982,Thorne:1986iy}. A static observer 
outside the horizon can replace the interior of a perturbed BH (left) with a perturbed fictitious membrane (right). The membrane is made of a viscous fluid whose properties (density, pressure, viscosity) are fixed to reproduce 
the BH phenomenology, in particular the QNM spectrum. In this paper we extend this concept to a generic dark compact object, 
parametrizing its properties in terms of the membrane's physical quantities.} 
\label{fig:paradigm}
\end{figure}

A dark compact object can be identified by two parameters, quantifying its shades of ``darkness'' and its 
``compactness''~\cite{Cardoso:2019rvt}. The darkness of the object depends on its reflectivity, whereas the 
compactness is related to the gravitational redshift at its surface.
It has been shown that the prompt ringdown signal which follows the merger is universal if the remnant is 
sufficiently compact~\cite{Cardoso:2016rao,Cardoso:2016oxy,Cardoso:2017cqb}. Putative near-horizon structures, which 
can be modelled in terms of a reflectivity coefficient at the surface of the 
object~\cite{Mark:2017dnq,Maggio:2017ivp}, 
show up only at late times in the form of repeated and modulated ``GW 
echoes''~\cite{Cardoso:2016rao,Cardoso:2016oxy,Abedi:2016hgu,Cardoso:2017cqb}. The time delay between echoes depends on the compactness of the object, whereas the relative amplitude between subsequent echoes is related to the reflectivity.
Considerable attention has been devoted to the study of this signal in a variety of models of near-horizon 
quantum 
structures~\cite{Conklin:2017lwb,Oshita:2018fqu,Wang:2019rcf,Cardoso:2019apo,Coates:2019bun,Buoninfante:2020tfb}, in the context of exotic 
compact objects in General Relativity~\cite{Ferrari:2000sr,Pani:2018flj}, and for modified-gravity 
scenarios~\cite{Holdom:2016nek,Buoninfante:2019swn,Buoninfante:2019teo,Delhom:2019btt}. Phenomenological waveform models 
have also been analyzed in 
detail~\cite{Nakano:2017fvh,Mark:2017dnq,Bueno:2017hyj,
Wang:2018mlp,Correia:2018apm,Wang:2018gin,Uchikata:2019frs,Testa:2018bzd,Maggio:2019zyv} and implemented in actual 
searches in 
LIGO/Virgo data~\cite{Abedi:2016hgu,Ashton:2016xff,Abedi:2017isz,Conklin:2017lwb,Westerweck:2017hus,Abedi:2018pst,
Conklin:2019fcs, 
Tsang:2019zra,Uchikata:2019frs} (see Refs.~\cite{Cardoso:2019rvt,Abedi:2020ujo} for some reviews).

The above studies focused on the case of objects almost as compact as a BH, but with some nonvanishing reflectivity. However, in what can be arguably considered as the best-motivated attempts to solve the information-loss paradox, the reflectivity is expected to be very small, for example owing to the enormous number of degrees of freedom that can be excited within 
a fuzzball~\cite{Guo:2017jmi} or due to nonlocal effects~\cite{Buoninfante:2019swn}. These solutions can be less compact than BHs; therefore, they could emit a ringdown signal which differs from the BH ringdown at early stages.

In this paper, we propose a generic and (as much as possible) model-independent framework to study the ringdown of a dark compact object in the context of General Relativity or some of its high-energy extensions.
We unveil some universal features of the ringdown and show that current GW observations can already place very stringent constraints on the compactness of a dark object.

%%%%%%%%%%%%%%%%%%%%%%%%%%%%%%%%%%%%%%%%
\section{Membrane paradigm for perturbed compact objects}\label{sec:framework}
%%%%%%%%%%%%%%%%%%%%%%%%%%%%%%%%%%%%%%%%
%%%%%%%%%%%%%%%%%%
\subsection{Setup}
%%%%%%%%%%%%%%%%%%
Our framework extends the seminal work by Damour, MacDonald, Price, Thorne, Znajek, and others on the BH 
membrane paradigm~\cite{Damour:1982,Thorne:1986iy,Price:1986yy}. In its standard formulation (see 
Fig.~\ref{fig:paradigm}), a static observer can replace the BH interior with a \emph{fictitious} 
membrane located at the horizon. The features of the internal spacetime are projected onto the 
membrane, whose physical properties are fixed by the Israel-Darmois junction 
conditions~\cite{Darmois1927,Israel:1966rt,VisserBook} (we assume $G=c=1$ units throughout), 
%%%%
\begin{equation}
 [[h_{ab}]]=0\,,\qquad [[K_{ab}-h_{ab} K]] = -8\pi T_{ab}\,, \label{junction}
\end{equation}
where $h_{ab}$ is the induced metric on the membrane, $K_{ab}$ is the extrinsic curvature, $K=K_{ab}h^{ab}$, $T_{ab}$ 
is the stress-energy tensor of the matter distribution located on the membrane, and $[[...]]$ denotes 
the jump of a quantity across the membrane.
%%%%
The junction conditions impose that the fictitious membrane is described by 
a viscous fluid whose thermodynamical properties (density, pressure, viscosity parameters) are uniquely determined if 
the membrane is demanded to act as a BH in terms of observable effects~\cite{Thorne:1986iy,Jacobson:2011dz}. 
We work in the framework of General Relativity\footnote{Although our framework can be applied to any 
theory of gravity, in practice, we assume that General Relativity works sufficiently well near the radius of the 
compact object. This assumption is well justified also for solutions 
to modified gravity in which putative extra degrees of freedom are heavy. In this case, all corrections to the metric 
and field equations are suppressed by powers of $\ell_P/R\ll1$, where $R$ is the object radius, and $\ell_P$ is the 
Planck length or the scale of new physics. For compact astrophysical objects, such corrections are negligible. No 
specific theory is assumed to describe the object interior. 
} and, for simplicity, we assume spherical symmetry. 

% %%%%%%%%%%%%%%%%%%%%%%%%%%%%%
\subsection{Background geometry}
% %%%%%%%%%%%%%%%%%%%%%%%%%%%%%

Let us assume that the fictitious membrane is located at the surface of the compact object, $r=R$, in 
some coordinate system. Owing to Birkhoff's theorem, the 
spacetime geometry for $r>R$ is described by the Schwarzschild metric,
\begin{equation}
{\rm d}s^2 = -f(r) {\rm d}t^2 + \frac{1}{f(r)} {\rm d}r^2 + r^2 \left({\rm d}\theta^2 + \sin^2{\theta} {\rm d}\varphi^2 
\right) \,, \label{backg-metric}
\end{equation}
where $f(r)=1-2M/r$ and $M$ is the total mass of the object. It is convenient to parametrize the circumferential radius 
of the membrane as 
%%%
\begin{equation}
 R=2M(1+\epsilon) \,,\label{radius}
\end{equation}
which defines the compactness of the object, $M/R$.
When $\epsilon\to0$ the membrane location coincides with the Schwarzschild's horizon, $R=2M$, whereas when 
$\epsilon=1/2$ it coincides with the light ring (or photon-sphere), $R=3M$, where circular, unstable photon orbits 
reside. As we shall show, in the latter limit some remarkable and universal properties of the QNM spectrum emerge. The 
case of $\epsilon \ll 1$ corresponds to ultracompact configurations characterized, for instance, by microscopic 
corrections at the horizon scale, whereas the case of $\epsilon\sim {\cal O}(0.1-1)$ corresponds to less compact objects 
whose compactness is comparable to or larger than that of a neutron star. 
In the context of fuzzballs, the former and latter 
cases describe tight and diffuse fuzzball models~\cite{Guo:2017jmi}, respectively. The model of 
nonlocal stars introduced in~\cite{Buoninfante:2019swn} is characterized by a maximal compactness of the order of the 
Buchdahl limit~\cite{Buchdahl:1959zz}, i.e., $\epsilon\simeq 0.125$, and therefore belongs to the latter category.

Following the original membrane paradigm, we assume that the fictitious membrane is such that the 
extrinsic curvature of the internal spacetime vanishes, $K_{ab}^-=0$~\cite{Thorne:1986iy}.
The details of the computation are given in Appendix~\ref{app:paradigm}, here we only summarize the main results. The 
junction conditions are compatible with a membrane described by the stress-energy 
tensor of a dissipative fluid~\cite{Thorne:1986iy,Oshita:2019sat}:
\begin{equation}
T_{ab}=\rho u_au_b+\left(p-\zeta \Theta \right)\gamma_{ab}-2\eta \sigma_{ab}\,,\label{stress-energy}
\end{equation}
where $\rho$, $p$, and $u_a$ are the density, the pressure, and the $3$-velocity of the fluid, respectively, whereas 
$\Theta$ is the expansion, $\sigma_{ab}$ is the shear tensor, and $\gamma_{ab}=h_{ab}+u_a u_b$ is the projector tensor. 
The parameters $\zeta$ and $\eta$ are the bulk and the shear viscosities that govern the response of the fluid to 
external perturbations. For a given model of the interior of the compact object, $\eta$ and $\zeta$ are uniquely 
determined. In general, they depend on the density (and the temperature) of the fluid and can be frequency-dependent, 
complex numbers. Since energy dissipation is absent when $\Re[\eta]=0=\Re[\zeta],$ for simplicity in the following we 
shall consider $\eta$ and $\zeta$ to be real constants, although our framework can be straightforwardly generalized.

When the object is unperturbed, the viscosity is irrelevant and the spherical membrane is described by a perfect 
fluid. The junction conditions then give the pressure and density of the fluid
%%%
\begin{equation}
 p(R)= \frac{2f(R)+Rf^{\prime}(R)}{16\pi R\sqrt{f(R)}}\,, \qquad \rho(R) = -\frac{\sqrt{f(R)}}{4\pi R}\,,
\end{equation}
%%%
which parametrically fix a barotropic equation of state, $p=p(\rho)$. In the BH limit, the 
density vanishes and the pressure diverges as the redshift factor $\epsilon^{-1/2}$. The speed of sound, 
$c_s=\sqrt{\partial p/\partial \rho}$,
% (see Eq.~\eqref{sound-speed} in Appendix~\ref{app:paradigm}), 
diverges both in the BH limit ($\epsilon\to0$) and in the light-ring limit 
($\epsilon\to1/2$). However, it is important to stress that these pathologies of the fluid are not an issue since the 
membrane is fictitious.

%%%%%%%%%%%%%%%%%%%%%%%%%%%%%%%%%%%%%%%%%%%%%
\subsection{Linear perturbations \& boundary conditions}
%%%%%%%%%%%%%%%%%%%%%%%%%%%%%%%%%%%%%%%%%%%%%

Gravitational perturbations in the exterior Schwarzschild geometry are governed by two simple 
equations that can be both written in a Schr\"{o}dinger-like 
form~\cite{Regge:1957td,Zerilli:1970se,Zerilli:1971wd,ChandraBook},
\begin{equation}
\frac{{\rm d}^2\psi(x)}{{\rm d}x^2}+\left[\omega^2-V(r)\right]\psi(x)=0 \,,\label{ODE}
\end{equation}
where the effective potential $V(r)$ reads
\begin{eqnarray}
\!\!\!\!\!V_{\rm axial}\! &=&\! f \left( \frac{l(l+1)}{r^2} - \frac{6M}{r^3} \right), \label{RW} \\
\!\!\!\!\!V_{\rm polar}\! &=&\! 2f \left(\! \frac{q^2(q+1)r^3 + 3q^2Mr^2 + 9M^2(qr+ M)}{r^3 (qr+3M)^2} \!\right)  
, \,\,\,\, \label{Zerilli}
\end{eqnarray}
for axial (i.e., magnetic or odd-parity) and polar (i.e., electric or even-parity) perturbations, respectively. Here 
$x$ is the ``tortoise'' coordinate such that ${\rm d}x/{\rm d}r=1/f(r)$, $q=\frac{1}{2}(l-1)(l+2)$, and $l$ is the 
angular number of the perturbation.

By imposing boundary conditions at infinity and at the inner boundary, 
Eq.~\eqref{ODE} defines an eigenvalue problem whose complex\footnote{In our conventions a stable mode corresponds to 
$\omega_I<0$, whereas an unstable mode corresponds to $\omega_I>0$.} eigenvalues are the QNMs of the system, 
$\omega=\omega_R+i\omega_I$. 
The effective potential vanishes at infinity, so the general solution to Eq.~\eqref{ODE} in the asymptotic 
region is a superposition of ingoing and outgoing waves. The QNMs are defined by imposing purely outgoing waves at 
infinity
\begin{equation}
\psi(x) \sim e^{i \omega x}\,, \qquad x \rightarrow \infty\,. \label{BC-infty}
\end{equation}

In the BH case, regularity at the horizon (located at $x\to-\infty$ in tortoise coordinates) implies purely 
ingoing waves at the inner boundary,
\begin{equation}
\psi_{\rm BH}(x) \sim e^{-i \omega x}\,, \qquad x \rightarrow -\infty\,, \label{BC-BH}
\end{equation}
for both axial and polar perturbations.

In the general case, the condition on the inner boundary depends on the properties of the object, in particular on its 
compactness and reflectivity. In order to derive these boundary conditions generically (i.e., without assuming any 
specific solution for the interior of the object), we rely on the membrane 
paradigm and impose junction conditions on the perturbations at the membrane. As detailed in 
Appendix~\ref{app:paradigm}, the generic boundary condition for the axial case turns out to be
%%%
\begin{equation}
 \frac{\psi'(x)}{\psi(x)} = -\frac{i\omega}{16\pi \eta} - \frac{R^2}{2(R-3M)}V_{\rm axial}(R)\,, 
\quad 
x\to x_R \label{BC-axial}
\end{equation}
%%%
where the prime denotes the derivative with respect to the tortoise coordinate, and $x_R \equiv x(R)$. The polar case is more involved but a similar computation yields 
%%%
\begin{equation}
 \frac{\psi'(x)}{\psi(x)} = -{16\pi i \eta\omega} + G(R,\omega,\eta,\zeta)\,, \qquad 
x\to x_R \label{BC-polar}
\end{equation}
%%%
where $G$ is a complicated function given in Appendix~\ref{app:paradigm}.
%%%

The boundary conditions given by Eq.~\eqref{BC-axial} and Eq.~\eqref{BC-polar} are one of our main results: they are valid for \emph{any} compact object described by a Schwarzschild exterior and have far-reaching consequences, which 
we discuss in the following.

The linear response of the object in the time 
domain is governed by the partial differential equation $[\partial_t^2-\partial_x^2 + V]\hat\psi(t,x)=0$, where 
$\hat\psi(t,x)$ is the inverse Fourier transform of $\psi(\omega,x)$.
In the axial sector, the time-domain response can be computed straightforwardly, since the inverse Fourier transform of 
the 
boundary condition~\eqref{BC-axial} reads
\begin{equation}
 \partial_x \hat\psi(t,x_R) = \frac{1}{16\pi\eta}\partial_t \hat\psi(t,x_R) - \frac{R^2 V_{\rm axial}(R)}{2(R-3M)} 
\hat\psi(t,x_R) \,.\label{BC-axialTD}
\end{equation}
The polar sector is more involved since the dependence on $\omega$ in Eq.~\eqref{BC-polar} is complicated and the 
inverse Fourier transform cannot be obtained in closed form.

\begin{figure*}[th]
\centering
\includegraphics[width=0.497\textwidth]{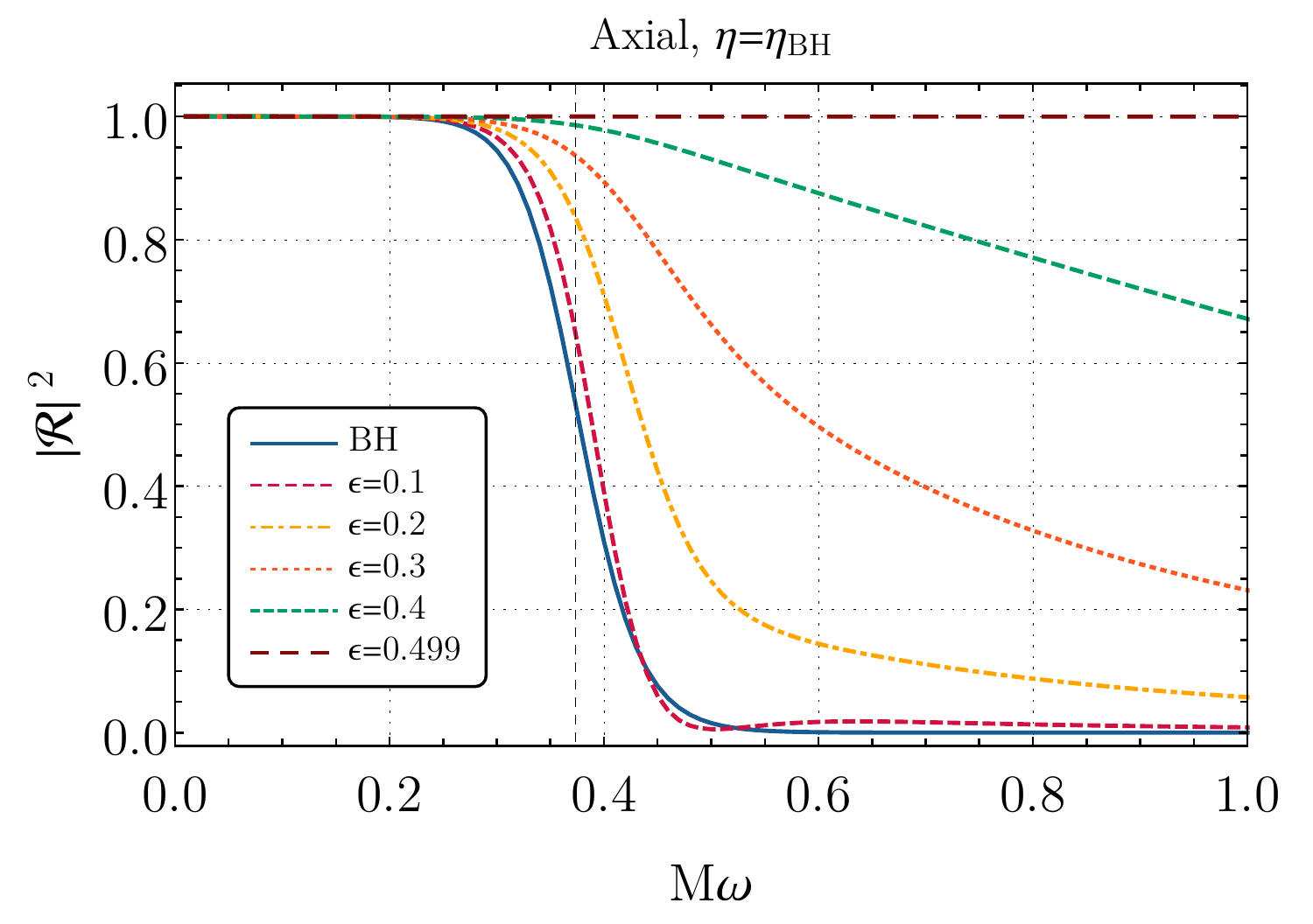}
\includegraphics[width=0.497\textwidth]{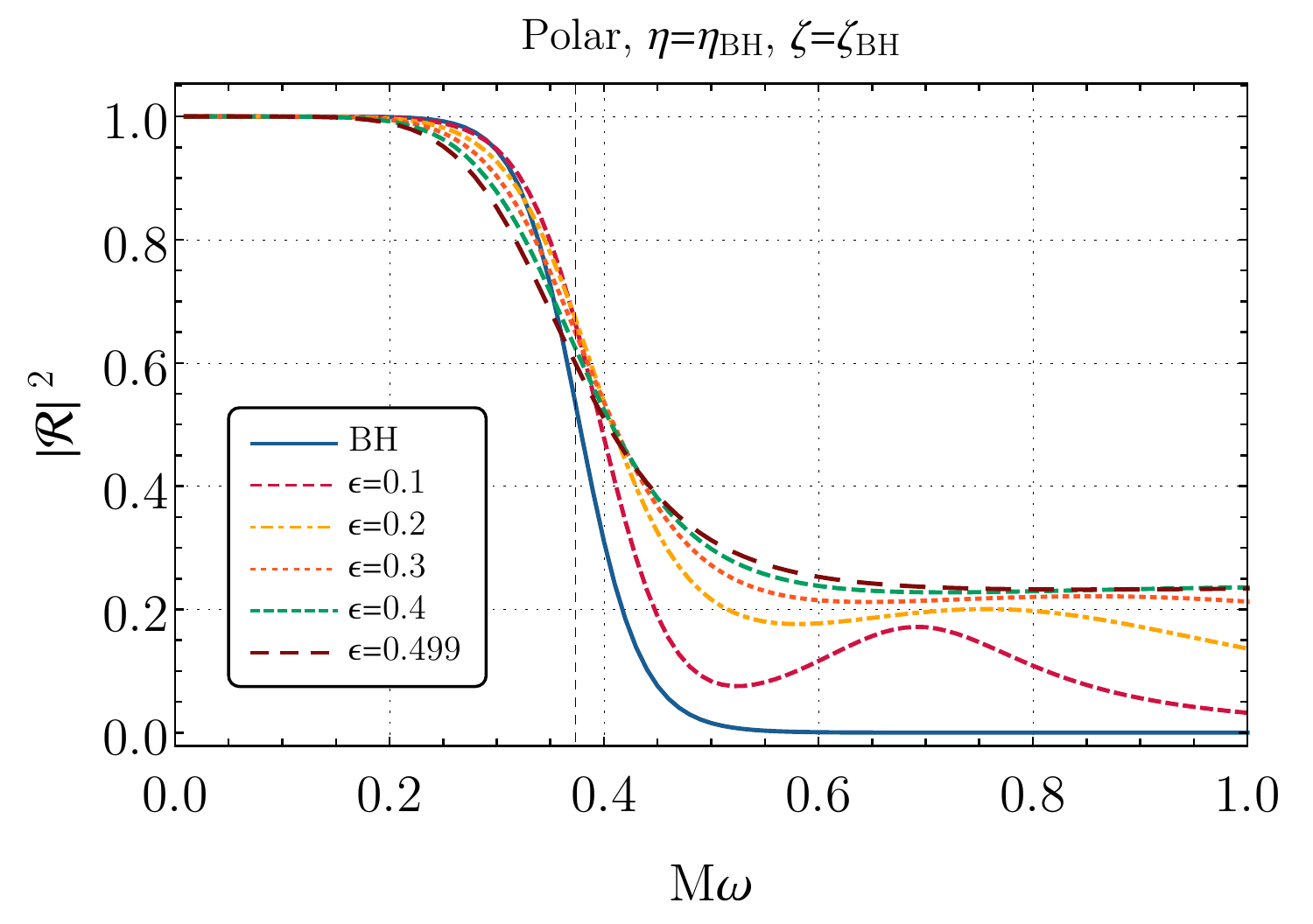}
\caption{Reflectivity coefficient of a dark compact object for axial (left panel) and polar (right panel) perturbations
in terms of the shear ($\eta=\eta_{\rm BH}$) and the bulk ($\zeta=\zeta_{\rm BH}$) viscosities of a fictitious fluid 
located at the surface of the object, $R=2M(1+\epsilon)$, and for several values of $\epsilon$. 
 At high frequency, $M\omega \gg1$, the object is perfectly 
absorbing for any value of $\epsilon$ [see Eq.~\eqref{R2}]. Interestingly, for $\epsilon \rightarrow 1/2$ the object 
becomes a perfect reflector of axial GWs for all frequencies and regardless of the values of $\eta$ (not shown in the 
plot). 
In both panels the vertical dashed line corresponds to the fundamental QNM frequency of a Schwarzschild BH, 
$M\omega_R^{\rm BH}\approx 0.3737$. 
}  
\label{fig:reflectivity}
\end{figure*}
%

%%%%%%%%%%%%%%%%%%%%%%%%%%%%%%%%%%%%%%%%%%%%%
\subsection{Effective viscosity, reflectivity, and universality}
%%%%%%%%%%%%%%%%%%%%%%%%%%%%%%%%%%%%%%%%%%%%%

In the BH limit the boundary conditions in Eqs.~\eqref{BC-axial} and~\eqref{BC-polar} must reduce to the BH one. 
A straightforward inspection of the above formulas shows that, when $R\to 2M$, the BH boundary 
condition in Eq.~\eqref{BC-BH} is recovered only if
%%%
\begin{equation}
 \eta  = \frac{1}{16\pi}\equiv \eta_{\rm BH}\,. \label{etaBH}
\end{equation}
%%%
In this case Eq.~\eqref{BC-axialTD} reduces to $\partial_x \psi(t,x) = \partial_t 
\psi(t,x)$, i.e., the standard boundary condition for an ingoing wave $\sim e^{-i\omega(t+x)}$ near the horizon.
Thus, a perturbed BH of mass $M$ behaves as a $(2+1)$-dimensional viscous fluid located at $r=R\sim 2M$ with 
shear viscosity $\eta=1/(16\pi)$. This result agrees with the standard BH membrane paradigm and extends the results of 
Refs.~\cite{Oshita:2019sat,Abedi:2020ujo} to the polar sector.

Unlike the axial sector, the polar boundary condition depends explicitly on the 
bulk viscosity $\zeta$. However, when $R\to2M$ the dependence on $\zeta$ disappears, so this 
parameter is not fixed by the linear perturbation analysis in the BH limit (see Appendix~\ref{app:paradigm} for details). 
On the other hand, according to the membrane paradigm, a generic BH can be mapped in terms of a fluid also endowed with 
a \emph{negative} bulk viscosity, $\zeta=-1/(16\pi)\equiv\zeta_{\rm BH}$. This peculiar property is associated with the 
\emph{teleological} nature of the event horizon~\cite{Thorne:1986iy}. 

One might wonder about the physical meaning of the shear viscosity as given in Eq.~\eqref{etaBH}. For this purpose, it 
is instructive to compute the 
reflectivity coefficient $|{\cal R}|^2$ of the object. The latter is defined through the scattering of a 
wave coming from infinity with (say) unitary amplitude and being partly reflected back, i.e.
%%%%
\begin{equation}
 \psi(x)\sim {\rm \cal R} e^{i\omega x} + e^{-i\omega x}\,, \quad x\to\infty \,, \label{scattering}
\end{equation}
%%%%
after being subjected to the boundary conditions~\eqref{BC-axial} 
or~\eqref{BC-polar} at $r=R$.

It is straightforward to compute ${\cal R}$ in the large-frequency\footnote{For an object with $M=10M_\odot$, radio 
waves with frequency $\gg 3\,{\rm kHz}$ are already in the large-frequency regime.} limit ($M\omega \gg1$), 
where the effective 
potential in Eq.~\eqref{ODE} can be neglected. In this case, we obtain a simple analytical result valid for both axial 
and polar perturbations:
%%%%
\begin{equation}
 |{\cal R}|^2=\left(\frac{1-\eta/\eta_{\rm BH}}{1+\eta/\eta_{\rm BH}}\right)^2\,. \label{R2}
\end{equation}
%%%%
This shows that the object is a \emph{perfect absorber} ($|{\cal R}|^2=0$) of high-frequency waves if and only if 
$\eta=\eta_{\rm BH}$, whereas it 
becomes a perfect reflector ($|{\cal R}|^2=1$) of high-frequency waves when either $\eta=0$ or $\eta\to\infty$. In the 
case of an ultracompact object with $\epsilon \ll 1$ and for any frequency, the former and latter cases correspond to Dirichlet 
($\psi=0$) and Neumann ($\psi'=0$) boundary conditions, respectively, for the axial 
sector. As clear from Eq.~\eqref{BC-polar}, the 
opposite is true for the polar sector: $\eta=0$ ($\eta\to\infty$) corresponds to the Neumann (Dirichlet) boundary 
condition.
These perfectly reflecting boundary conditions were studied in 
Refs.~\cite{Maggio:2017ivp,Maggio:2018ivz}, of which our results are a generalization.

Although $\eta$ is formally a free parameter, we expect the most interesting range to be $\eta\in[0,\eta_{\rm BH}]$. 
Indeed, from Eq.~\eqref{R2} negative values of $\eta$ would correspond to $|{\cal R}|^2>1$, which leads to 
superradiant instabilities~\cite{Brito:2015oca} even in the static case. Similarly, for 
$\eta>\eta_{\rm BH}$ the reflectivity coefficient is a monotonically growing function of the shear viscosity, which is 
also unphysical.

The reflectivity coefficient for generic frequencies can be computed numerically with standard methods (see e.g., 
Ref.~\cite{Brito:2015oca}). The result is shown in Fig.~\ref{fig:reflectivity} for axial (left panel) and polar (right 
panel) perturbations and some representative choices of the parameters. Besides recovering all the expected 
limits with respect to the BH case and in the large-frequency regime, this plot also shows some other interesting 
effects.

Noticeably, as $\epsilon\to1/2$ 
(i.e., as the surface of the object approaches the light ring), the reflectivity coefficient of axial GWs tends to 
unity for any frequency.
This can be understood by looking at the boundary condition~\eqref{BC-axial}, whose last term diverges as
$R\to3M$, thus imposing $\psi=0$ for any frequency and for any $\eta\in \mathbb{C}$. 
Therefore, regardless of its internal structure, an object with $R=3M$ is a \emph{perfect reflector} of axial 
GWs\footnote{\label{foot-except}The only exception is when $\eta\to-\frac{3i\omega}{16\pi q} (R-3M)$ as $R\to3M$, in 
which case the divergence in Eq.~\eqref{BC-axial} cancels out. However, for any purely imaginary $\eta$ the 
axial reflectivity is unity at any frequency, so the object remains a perfect reflector. 
Although very fine-tuned, this particular case corresponds to 
thin-shell gravastars~\cite{Visser:2003ge,Pani:2010em}.}. 
The polar sector does not enjoy the same universality. Moreover, in some regions of the parameter space the polar 
reflectivity shows some distinctive peak at specific frequencies, as shown in Fig.~\ref{fig:reflectivity} for 
$\epsilon=0.1$ (for the choice $\eta=\eta_{\rm BH}$ and $\zeta=0$ not shown in the plot the peak can be as high 
as $|{\cal R}|^2=1$).

It is also interesting to note that the reflectivity of the spacetime at intermediate frequencies 
($M\omega={\cal O}(0.1-1)$, those relevant in the ringdown~\cite{Abbott:2016blz}) can be larger than the BH 
reflectivity even if the object is a perfect absorber at higher frequencies. 
For example, at the fundamental BH QNM frequency, $M\omega \approx 0.3737$ (shown in Fig.~\ref{fig:reflectivity} as a 
vertical dashed line), the reflectivity is $|{\cal R}|^2\approx 1$ for axial perturbations, $\epsilon=0.4$ and 
$\eta=\eta_{\rm BH}$, although $|{\cal R}|^2\to0$ as $M\omega \gg1$. 

These effects have dramatic consequences for the ringdown and the QNM spectrum, as we are going to discuss in the next 
section.

%%%%%%%%%%%%%%%%%%%%%%%%%%%%%%%%%%%%%%%
\section{Ringdown \& QNM spectrum}\label{sec:RD}
%%%%%%%%%%%%%%%%%%%%%%%%%%%%%%%%%%%%%%%

Equation~\eqref{ODE} with boundary conditions~\eqref{BC-infty} and either~\eqref{BC-axial} (for the axial case) or 
\eqref{BC-polar} (for the polar case) can be solved numerically as an eigenvalue problem to derive the QNM spectrum of a 
given model, identified by the parameters $(\epsilon,\eta,\zeta)$. For each value of $l=2,3,...$, 
there exists a countably infinite set of QNMs which can be ordered ($n=0,1,2,...$) in terms of their imaginary part, 
$\Im(\omega)\equiv \omega_I$: the fundamental mode ($n=0$) has the longest decay time, 
$\tau=-1/\omega_I$, whereas its overtones ($n=1,2,...$) have a shorter lifetime. 

In the axial case the QNM spectrum depends on the two-dimensional parameter space $(\epsilon, \eta)$ --~i.e., on the 
compactness and viscosity of the membrane~-- whereas in the polar sector there is an additional dependence on the bulk 
viscosity $\zeta$. As previously discussed, in the $\epsilon\to0$ limit the polar boundary conditions (and hence the 
QNMs) are independent of $\zeta$.

We compute the QNM spectrum using two numerical methods~\cite{Pani:2013pma}: a shooting method based on the 
direct integration of Eq.~\eqref{ODE}, and another one based on continued fractions~\cite{Leaver:1985ax,Berti:2009kk} 
in a variant adapted from the case of compact stars~\cite{Leins:1993zz,Benhar:1998au,Pani:2009ss}. The continued-fraction method is 
more robust as it is well suited also for overtones with $M\omega_I \gg1$, for which the direct integration fails. When 
both applicable, the two methods are in excellent agreement.

%%%%%%%%%%%%%%%%%%%%%%%%%%%%%%%%%%%%%%%%%%%%%%%
\subsection{Prompt ringdown versus GW echoes}
%%%%%%%%%%%%%%%%%%%%%%%%%%%%%%%%%%%%%%%%%%%%%%%

Before proceeding with the QNM analysis, let us look at the linear response in the time domain by studying a 
wave packet scattered off the object. Since the external spacetime for $r>R$ is described by the Schwarzschild 
metric, the initial, \emph{prompt ringdown} is associated with the scattering of the wave packet off the effective 
potential $V(r)$ [see Eq.~\eqref{ODE}] independently of the boundary conditions. Indeed, if $\epsilon$ is sufficiently 
small the following causality argument shows that the 
boundary conditions cannot affect the prompt ringdown~\cite{Cardoso:2016rao}. The decay time 
scale of the prompt ringdown is associated with the instability time scale of circular photon orbits 
at the light ring~\cite{Cardoso:2014sna,Cardoso:2019rvt}, or equivalently to the decay time of the fundamental QNM of a 
Schwarzschild 
BH, $\tau=-1/\omega_I \approx 10 M$. Thus, the boundary conditions at $r=R~(<3M)$ do not have time to modify the prompt 
ringdown if the round-trip time of the radiation from the photon sphere to the boundary and 
back~\cite{Cardoso:2016oxy,Cardoso:2017cqb,Cardoso:2019rvt},
%%%
\begin{equation}
 \tau_{\rm echo} = 2 M \left[1 - 2 \epsilon - 2\log(2\epsilon)\right]\,, \label{tauecho}
\end{equation}
%%%
is much longer than $\approx 10 M$. This imposes $\epsilon\ll {\cal O}(0.01)$~\cite{Cardoso:2019rvt}. Objects with such a
large compactness were dubbed \emph{ClePhOs}, since they have a clean 
photon-sphere~\cite{Cardoso:2017cqb,Cardoso:2019rvt}.
On the other hand, if $\epsilon\gtrsim {\cal O}(0.01)$ the object's interior is expected to affect the prompt 
ringdown.

\begin{figure}[t]
\centering
\includegraphics[width=0.485\textwidth]{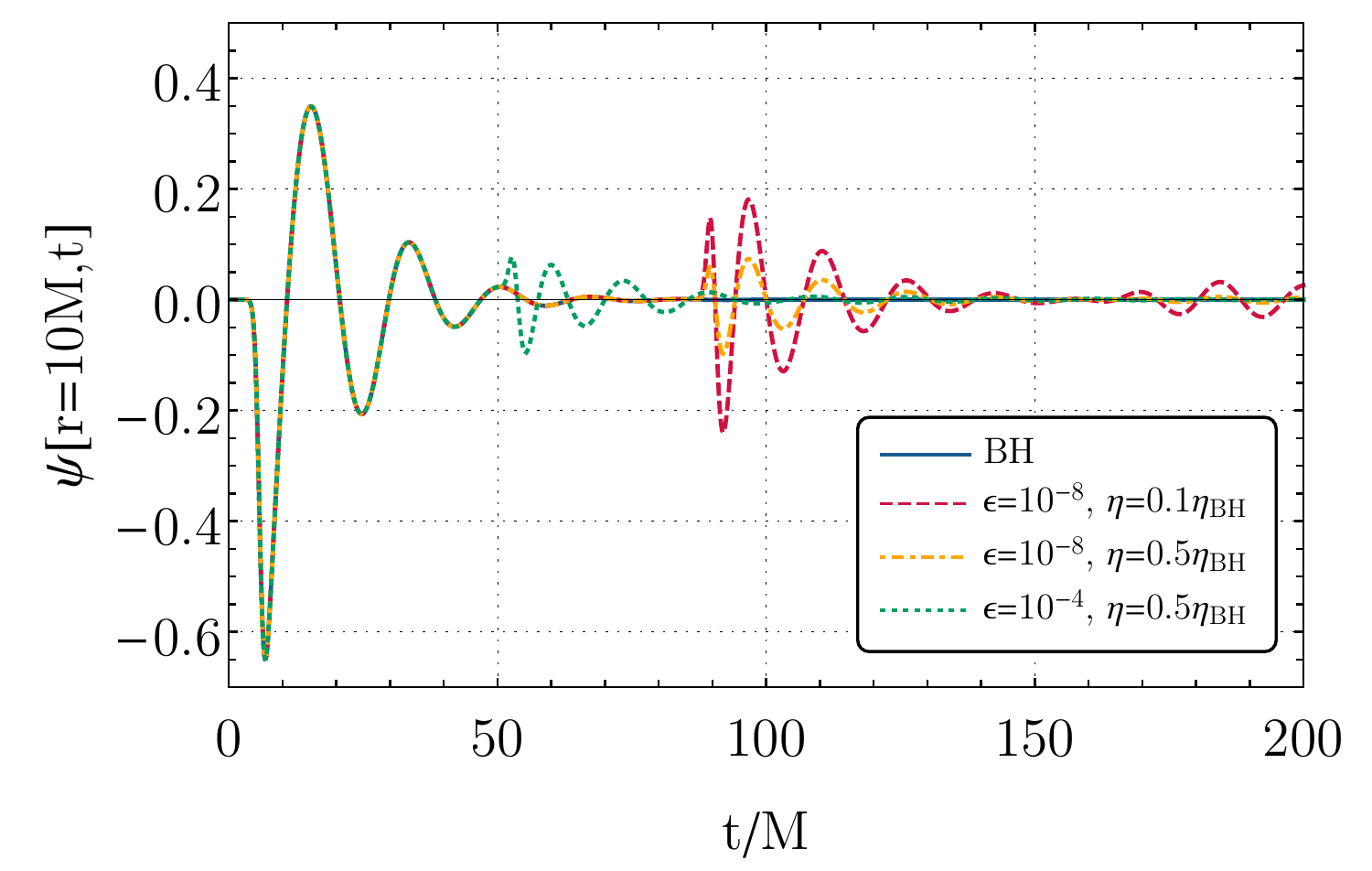}
\includegraphics[width=0.485\textwidth]{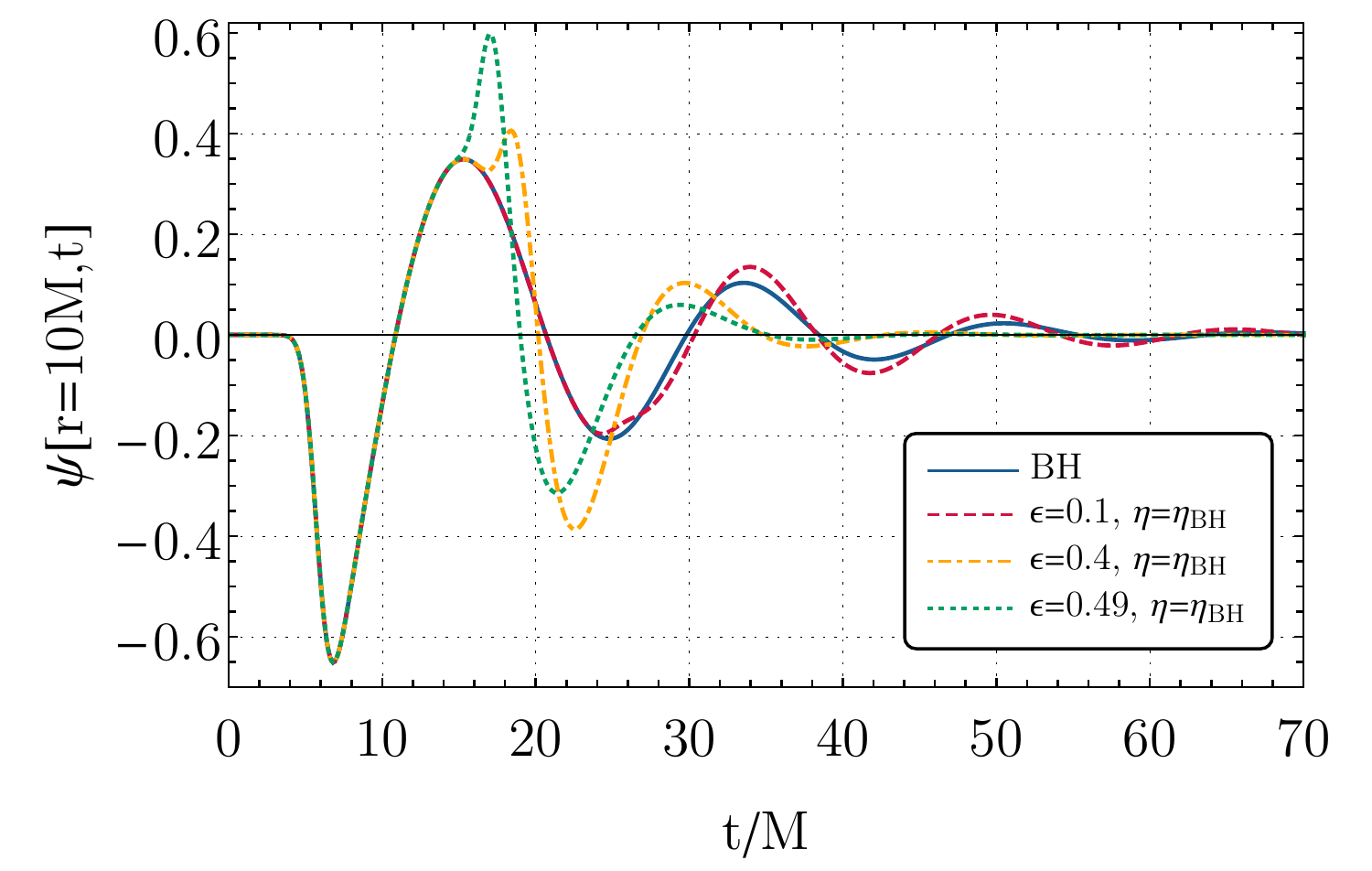}
\caption{
Ringdown of a dark compact object with radius $R=2M(1+\epsilon)$ and effective shear viscosity $\eta$. We consider 
axial perturbations and an initial Gaussian profile, $\psi(x,0)=0,\,\, \partial_t \psi(x,0)=\exp[{-(x-7)^2}]$. 
Top panel: when $\epsilon\ll0.01$ GW echoes appear for several values of $\eta$ related to the reflectivity of the 
object.
Bottom panel: a selection of ringdown waveforms for $\epsilon>0.01$. In this case echoes are absent but the prompt 
ringdown is governed by the modified QNMs of the object.
} 
\label{fig:ringdown}
\end{figure}
\begin{figure*}[th]
	\centering
	\includegraphics[width=0.49\textwidth]{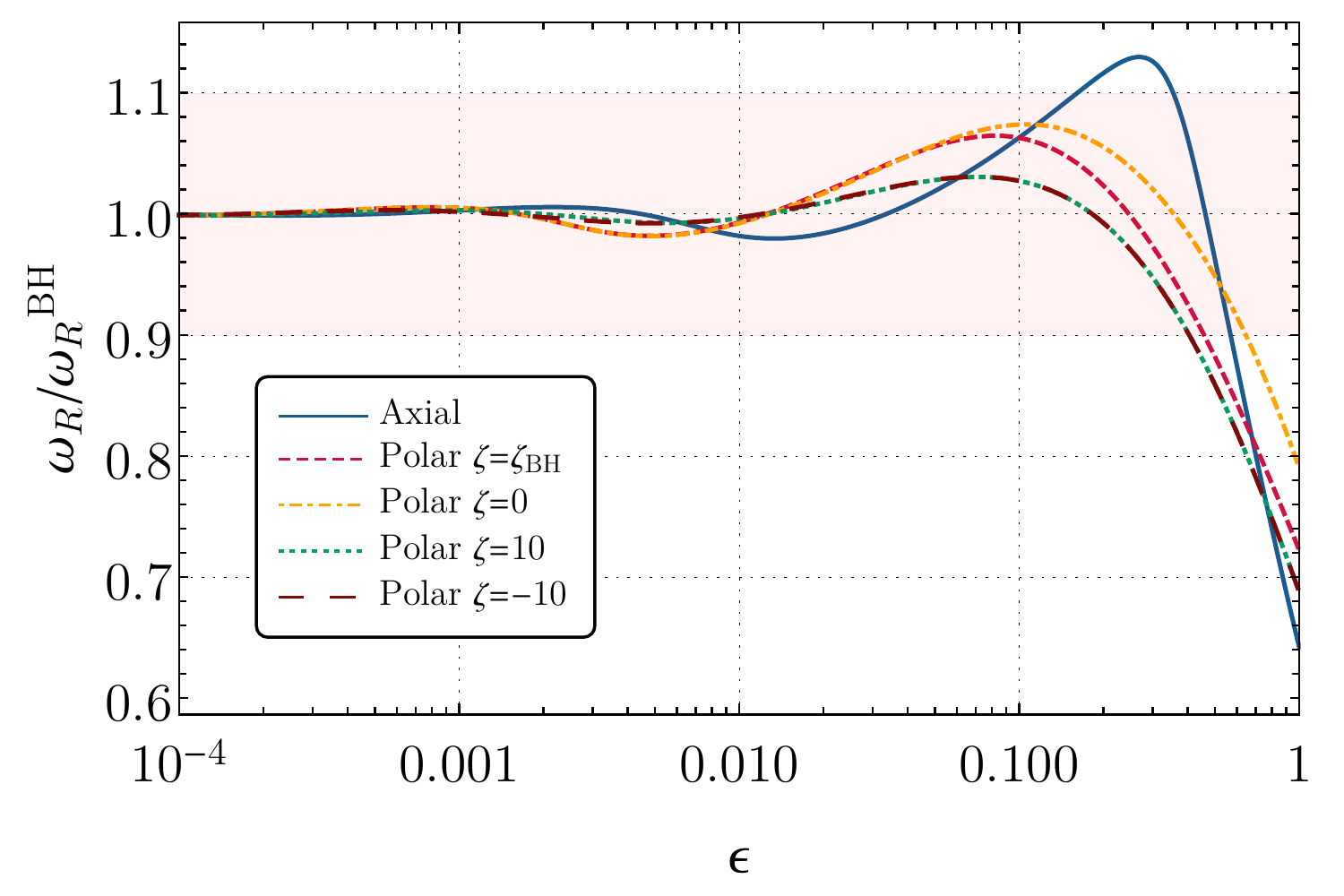}
	\includegraphics[width=0.49\textwidth]{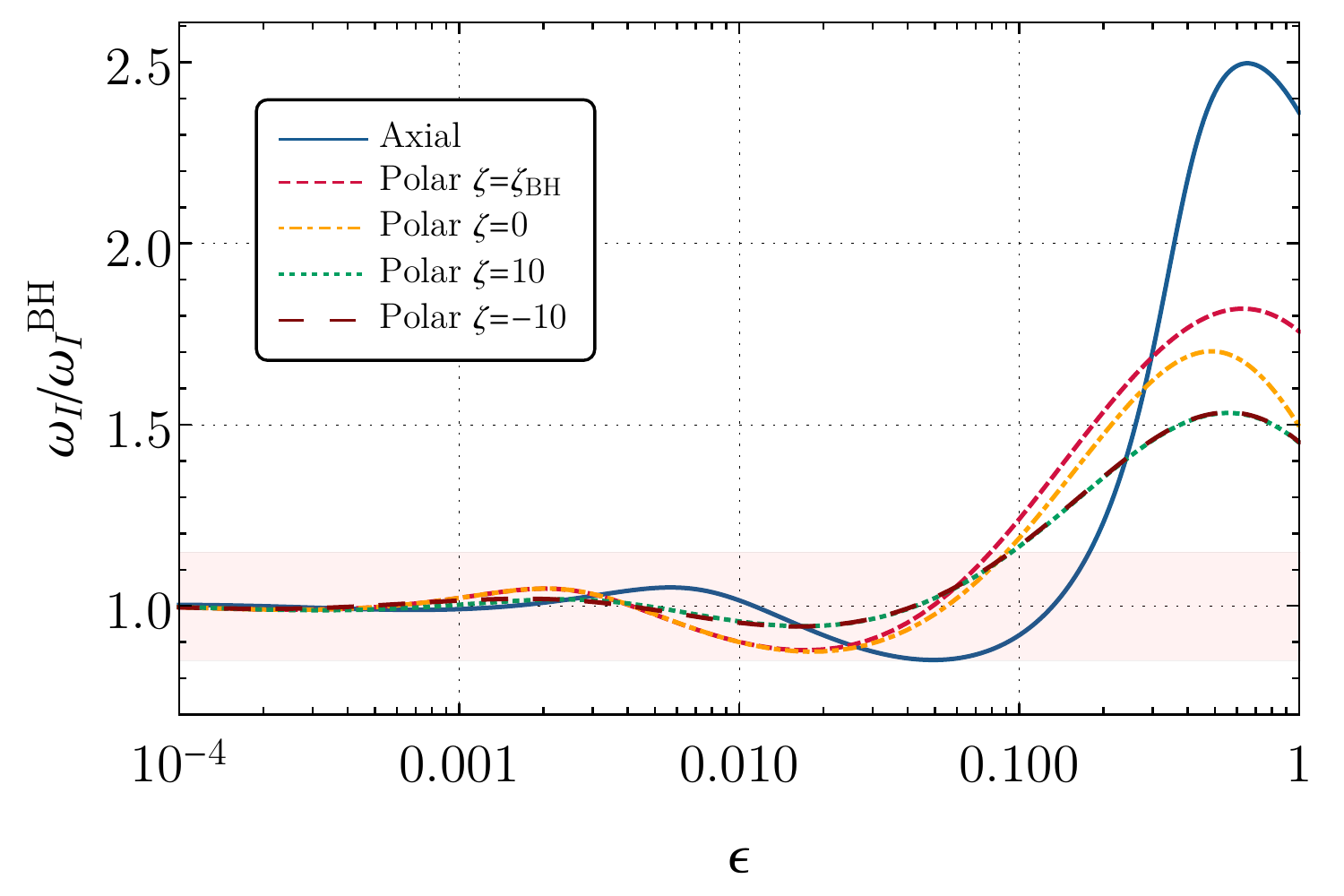}
\caption{Real (left panel) and imaginary (right panel) part of the fundamental ($l=2$) QNM of a dark compact 
object with effective shear viscosity $\eta=\eta_{\rm BH}$, with respect to their corresponding value in the BH 
case. The QNMs are functions of the parameter $\epsilon$ which is related to the compactness of the object. For 
$\epsilon \rightarrow 0$ we recover the fundamental BH QNM and axial and polar modes are isospectral, whereas for 
larger values of $\epsilon$ (less compact object) the QNMs differ from the BH case and isospectrality is broken. For 
large $\zeta$ or for $\epsilon\to0$, the polar QNMs are independent of $\zeta$. As a reference, the shaded regions correspond to the error bars ($10\%$ and $15\%$, respectively) on the 
frequency and the damping time for the merger event GW150914~\cite{Abbott:2016blz,TheLIGOScientific:2016src}. For 
$\epsilon\gtrsim 0.01$, these QNMs appear in the modified prompt ringdown, see e.g. Fig.~\ref{fig:ringdown}.
}
\label{fig:QNMeps}
\end{figure*}

These expectations are confirmed by the ringdown waveforms shown in Fig.~\ref{fig:ringdown}, which are obtained
by solving the linearized problem for axial perturbations with boundary condition~\eqref{BC-axialTD} using a 
fourth-order Runge-Kutta finite-difference scheme. 
The top panel shows the $\epsilon\to0$ limit. Confirming previous 
results~\cite{Cardoso:2016rao,Cardoso:2016oxy,Cardoso:2017cqb}, in this case the prompt ringdown 
is universal and the details of the object's interior appear only in late-time echoes of the initial ringdown. Their 
period is roughly given by Eq.~\eqref{tauecho}, and their frequency content and phase are modulated by the boundary 
conditions and by the tunneling through the potential barrier~\cite{Testa:2018bzd,Maggio:2019zyv}. Finally, their 
amplitude depends on 
the shear viscosity $\eta$. In particular, $\eta\approx0$ corresponds to $|{\cal R}|^2\approx1$ for which subsequent 
echoes amplitudes are only mildly damped~\cite{Testa:2018bzd}, whereas absorption is maximized as $\eta\to\eta_{\rm 
BH}$. 
In the latter case the linear response is identical to that of a BH, since the boundary conditions are the same when 
$\eta\to\eta_{\rm BH}$ and $\epsilon\to0$.

The bottom panel of Fig.~\ref{fig:ringdown} focuses on the case $\epsilon\gtrsim {\cal O}(0.01)$, where the 
prompt ringdown is modified and no subsequent echoes appear. The changes to the prompt ringdown can be understood by 
considering that the part of the wave packet that initially tunnels through the barrier has enough time to be reflected 
at $r=R$ and to tunnel again to infinity. This process results in a superposition of the two pulses (the one directly 
reflected 
by the potential barrier and the one reflected by the object), which interfere in a complicated pattern. When the two 
pulses sum in phase, the interference produces higher peaks in the prompt ringdown.
Crucially, as discussed in the following section, at late time the prompt ringdown is dominated by the fundamental QNM of the 
object, which is \emph{not} the mode of the universal prompt ringdown in the BH case.

Finally, one might wonder why there are no echoes for $\epsilon\gtrsim {\cal O}(0.01)$, even though the reflectivity 
of the object at intermediate frequencies is high, as shown in Fig.~\ref{fig:reflectivity}.
The reason is that only waves with frequency $V(R)<\omega^2<V_{\rm max}$ can be trapped between the object surface and 
the potential barrier. Therefore, when the compactness decreases, the resulting cavity is small. Furthermore, 
the transmission coefficient of the potential barrier is large when $\omega^2\lesssim V_{\rm max}$, which implies that 
these frequencies cannot be trapped efficiently. This also explains why when $R\to 3M$ (corresponding to perfect 
reflection 
of axial GWs) there are no echoes, since the membrane is very close to the maximum of the potential. In 
practice, for $\epsilon\gtrsim{\cal O}(0.01)$ one sees only the interference between the prompt 
ringdown and the first echo, while subsequent reflections are strongly suppressed or absent, as in the bottom 
panel of Fig.~\ref{fig:ringdown}.

%%%%%%%%%%%%%%%%%%%%%%%%%%%%%%%%%%%%%%%%%%%%%%%
\subsection{QNM dependence on the compactness and viscosity}
%%%%%%%%%%%%%%%%%%%%%%%%%%%%%%%%%%%%%%%%%%%%%%%
%

Let us first analyze the QNM spectrum of a dark compact object with $\eta=\eta_{\rm BH}$. Figure~\ref{fig:QNMeps} shows
the real and imaginary part of the axial and polar QNMs as a function of $\epsilon$, with respect to the BH case.
As expected, for $\epsilon \rightarrow 0$ we recover the fundamental $l=2$ QNM of a Schwarzschild BH, $M\omega_{\rm BH} 
\approx  0.3737 - i0.089$. For larger values of $\epsilon$, the compactness of the object decreases and the QNMs depart 
from the BH case. From the previous discussion, for $\epsilon\ll0.01$ the (small) deviations to the QNMs do not affect 
the prompt ringdown. Furthermore, for $\eta=\eta_{\rm BH}$ echoes are suppressed, so an object with $\epsilon \ll 0.01$ and 
$\eta=\eta_{\rm BH}$ is indistinguishable from a BH in terms of its ringdown.

For $\epsilon \gtrsim 0.01$, the prompt ringdown is affected by the modified QNMs. Indeed, by fitting the time-domain 
waveform at late times with a damped sinusoid, $\Psi(t)\sim A\cos(\omega_R 
t+\phi)e^{-t/\tau}$, we can verify that in this range the prompt ringdown is governed by
the \emph{modified} fundamental QNM of the system. This noteworthy feature carries a characteristic imprint of the 
structure of the remnant.

The QNMs of the Schwarzschild BH enjoy a remarkable ``isospectrality''~\cite{Chandrasekhar:1975zza,ChandraBook}: the 
entire spectrum is the same for the axial and the polar gravitational sectors. This isospectrality is broken for 
finite values of $\epsilon$ (and for $\eta\neq 
\eta_{\rm BH}$, as discussed below). For each point of the parameter space, the $l=2$ fundamental QNMs form a 
\emph{doublet}.

Polar modes are qualitatively similar to the axial ones and show a mild dependence on the bulk viscosity $\zeta$.
The QNM spectrum is independent of $\zeta$ in the large-$\zeta$ limit, as shown in Fig.~\ref{fig:QNMeps} by the 
$\zeta=\pm10$ curves. This is expected since, when $\zeta\to\infty$, the polar boundary condition~\eqref{BC-polar} is 
independent of $\zeta$, and the second term of the right-hand side reduces to $G=\frac{3 M (R-2 M)}{R^2 (3 M+q R)}$.

\begin{figure}[t]
\centering
\includegraphics[width=0.49\textwidth]{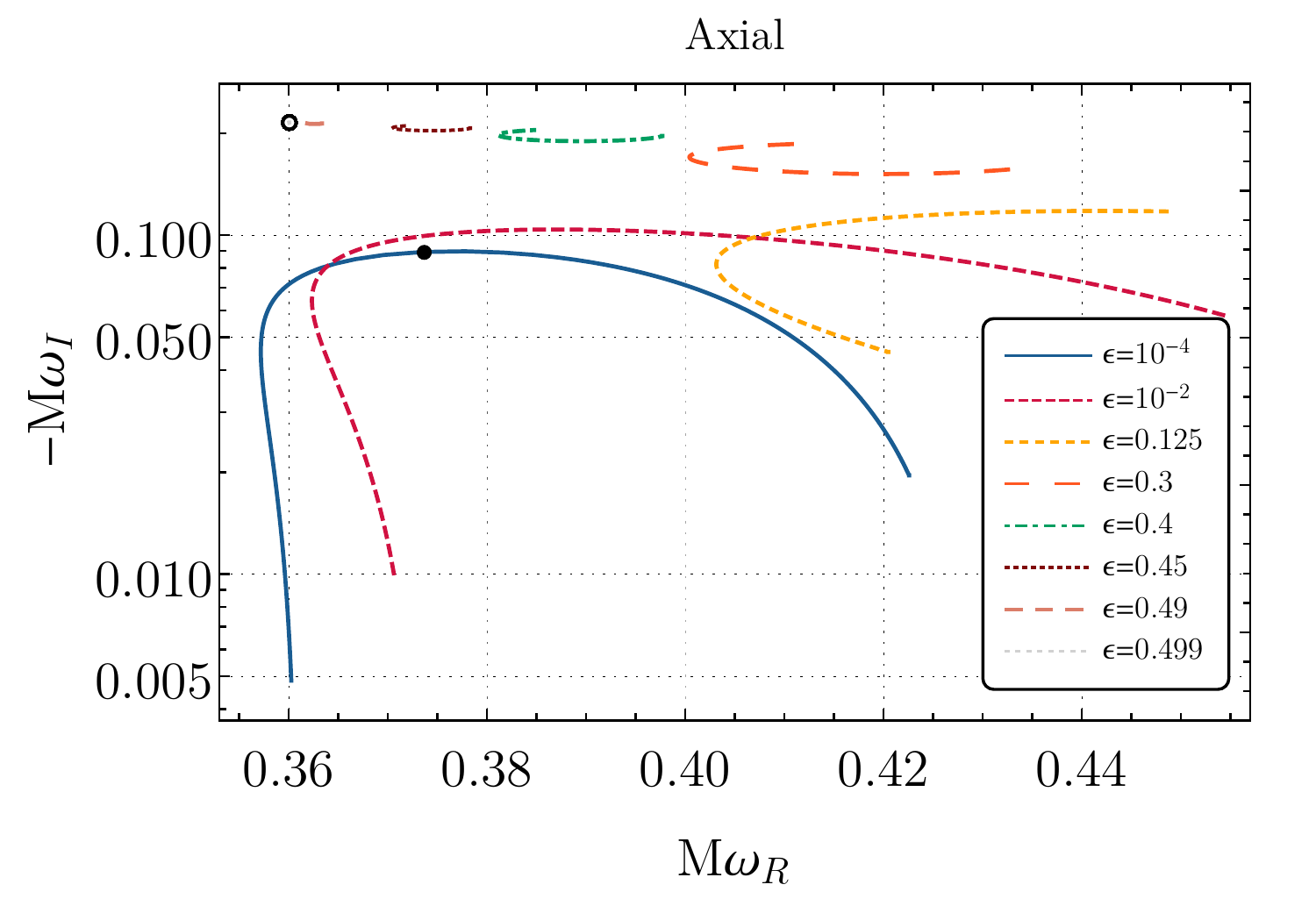}\\
%%%
\caption{
The complex QNM plane for axial modes. Each curve is parametrized in terms of the effective shear viscosity $\eta\in[10^{-4},100]$. As $\epsilon 
\rightarrow 1/2$, the curves collapse to a single \emph{universal} point (black empty circle), corresponding to the 
universal QNM in Eq.~\eqref{QNMuniversal}. As a reference, the fundamental QNM of a Schwarzschild BH is marked by a 
black dot. 
} 
\label{fig:QNMmanyeps}
\end{figure}

Finally, Fig.~\ref{fig:QNMmanyeps} shows the complex QNM plane of a dark compact object for axial perturbations. For 
a given value of $\epsilon$, each 
curve is parametrized by the shear viscosity\footnote{We focus on $\eta>0$ because, consistently with the previous 
analysis 
of the reflectivity coefficient, when $\eta<0$ we found \emph{unstable} modes (i.e., modes with $\omega_I>0$). This is 
expected since a negative shear viscosity corresponds to energy injected into the system, i.e., to a 
super-emitter with $|{\cal R}|^2>1$~\cite{Brito:2015oca}.} $\eta$.
As the location of the surface approaches the light ring 
($\epsilon\to1/2$), the axial QNMs become independent of $\eta$. Indeed, they tend to a universal mode which, for 
$l=2$, reads 
%%%
\begin{equation}
 M\omega_{\rm axial} \approx 0.3601 -i 0.2149\,, \quad \epsilon\to1/2\,, \label{QNMuniversal}
\end{equation}
%%%
for any value of $\eta$. As previously discussed, in this limit the object becomes a perfect reflector of axial 
GWs, regardless of the value of $\eta$. Therefore, a compact object with $R=3M$ has a \emph{universal axial QNM 
spectrum}, regardless of its properties (but see footnote~\ref{foot-except} for a loophole).
Unfortunately, this remarkable universality does not apply to the polar sector, since there does not exist a value of 
$R$ at which the function $G$ defined in Eq.~\eqref{BC-polar} diverges for any value of $\eta$, $\zeta$, and $\omega$.

%%%%%%%%%%%%%%%%%%%%%%%%%%%%%%%%%%%%%%%%%%%%%%%%%%
\subsection{Transition from fundamental modes to overtones \& instability of the QNM spectrum}
%%%%%%%%%%%%%%%%%%%%%%%%%%%%%%%%%%%%%%%%%%%%%%%%%%

As previously discussed, the computation of the QNMs in the $\epsilon\to0$ limit is less relevant from a phenomenological point 
of view, since these modes do not govern the prompt ringdown but only the late-time echo 
structure~\cite{Cardoso:2019rvt}. Nonetheless, from a theoretical standpoint it is enlightening to study their 
dependence on $\eta$, since this parameter interpolates between the BH limit ($\eta=\eta_{\rm BH}$) and the perfectly 
reflecting case ($\eta\to0,\infty$).

\begin{figure}[t]
\centering
\includegraphics[width=0.49\textwidth]{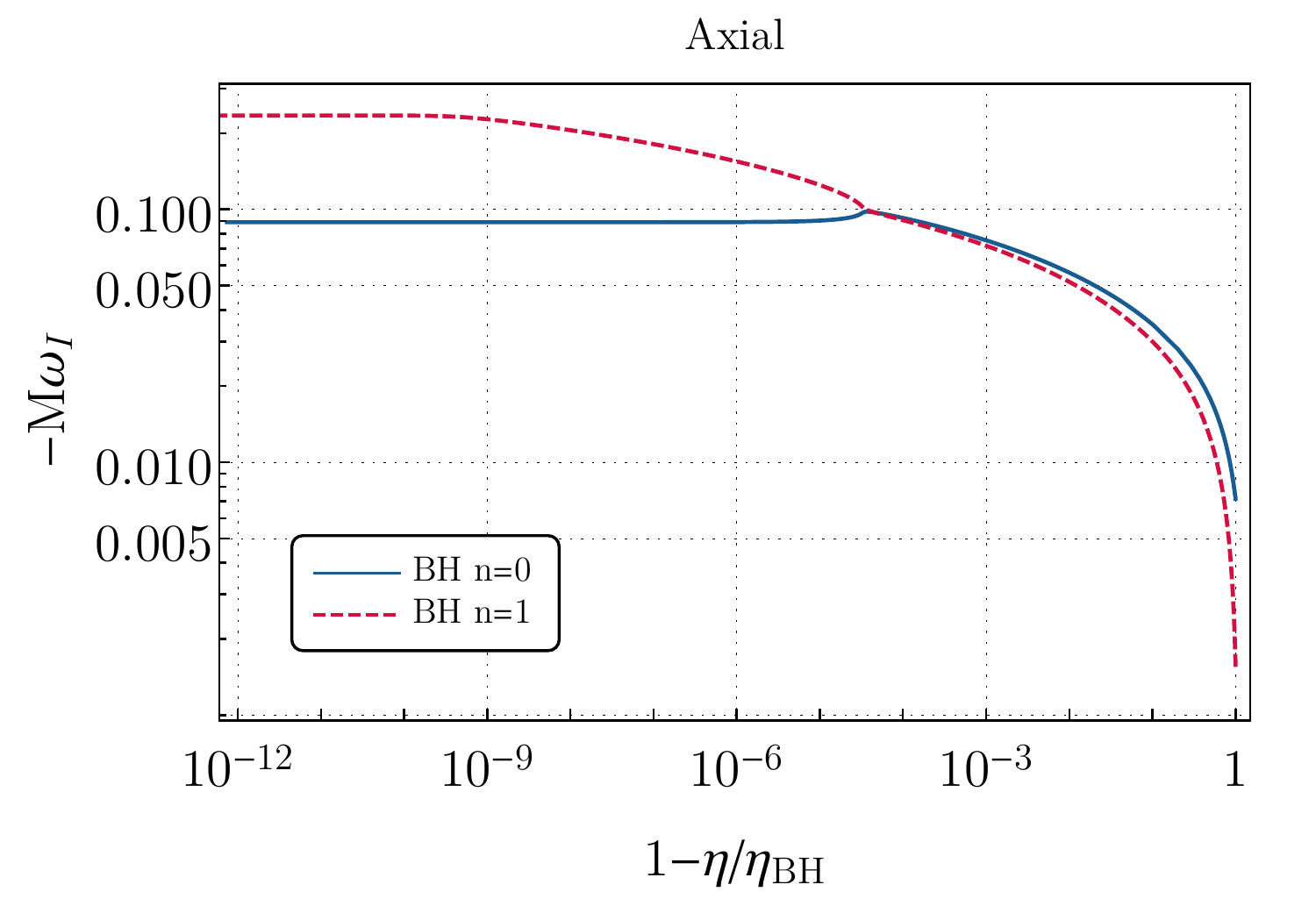}
\caption{Imaginary part of the axial QNMs of a dark compact object with radius $R=2M(1+\epsilon)$ and 
$\epsilon=10^{-10}$ as a function of the effective shear viscosity, tracking the fundamental ($n=0$, blue curve) and 
first overtone ($n=1$, red curve) QNM of a Schwarzschild BH (leftmost part of the plot) down to the limit of perfectly 
reflecting object ($\eta=0$, rightmost part). There exists a crossing point at which the imaginary part of the BH 
overtone becomes smaller than 
that of the BH fundamental mode. The polar case is qualitatively similar.} 
\label{fig:overtone}
\end{figure}

For concreteness, we consider $\epsilon=10^{-10}$ and track the QNMs from the BH case ($\eta=\eta_{\rm BH}$) down to 
$\eta=0$. In Fig.~\ref{fig:overtone}, we show the imaginary part of the axial QNMs obtained with this procedure 
starting from the fundamental ($n=0$, blue curve) mode and from its first overtone ($n=1$, red curve). As expected, at 
the starting point we recover the modes of a Schwarzschild BH. 
However, immediately after $\eta$ decreases (going on the right in the plot) there is a crossing point between the two 
curves: already when $\eta<0.99997\eta_{\rm BH}$ the first BH overtone becomes longer lived (i.e., smaller $\omega_I$) 
than the fundamental BH mode, showing that the former becomes more relevant for the 
late-time dynamics of the system\footnote{Recently, it has been realized that the overtones might play an important role 
in the ringdown of loud 
mergers~\cite{Giesler:2019uxc,Isi:2019aib,Berti:2007zu,Baibhav:2017jhs,Bhagwat:2019dtm,Ota:2019bzl,	
Forteza:2020hbw}. Studying the implication of our results for overtones is an interesting extension of this work.}. 
Although tracking higher overtones is challenging, we have indications that this 
behavior is rather general: higher overtones increasingly become more relevant in the $\eta\to0$ limit.

Indeed, neither of the modes shown in Fig.~\ref{fig:overtone} for $\eta\to0$ correspond to the fundamental mode of the perfectly 
reflecting case. The latter is a long-lived mode with a tiny imaginary part~\cite{Maggio:2018ivz}.
Interestingly enough, tracking the fundamental mode from the perfectly reflecting case back to $\eta_{\rm BH}$ is 
extremely challenging. We suspect that this is connected to the fact that --~when tracked to $\eta\to\eta_{\rm BH}$~-- 
this mode corresponds to a very high-order overtone in the BH limit, which are extremely difficult to compute 
numerically, due to their large imaginary part. 
Furthermore, an interesting property of the BH QNM spectrum is its instability against small deformations of the 
eigenvalue problem, such as bumps in the effective potential~\cite{Nollert:1996rf,Leung:1999iq,Barausse:2014tra} or 
slightly different boundary conditions~\cite{Cardoso:2016rao} (see Ref.~\cite{Jaramillo:2020tuu} for recent work on this 
topic). In our numerical analysis we have seen hints of this instability (which is more severe for high-order 
overtones~\cite{Jaramillo:2020tuu}) due to finite-$\epsilon$ effects and to the slightly different boundary conditions 
when approaching the BH limit. We defer a detailed analysis of this interesting problem to the future, stressing that 
our framework is well-suited to explore this problem in full generality.

%

%%%%%%%%%%%%%%%%%%%%%%%%%%%%%%%%%%%%%%%%%%%%%%%
\subsection{Detectability}
%%%%%%%%%%%%%%%%%%%%%%%%%%%%%%%%%%%%%%%%%%%%%%%

%
\begin{figure*}[th]
\centering
\includegraphics[width=0.48\textwidth]{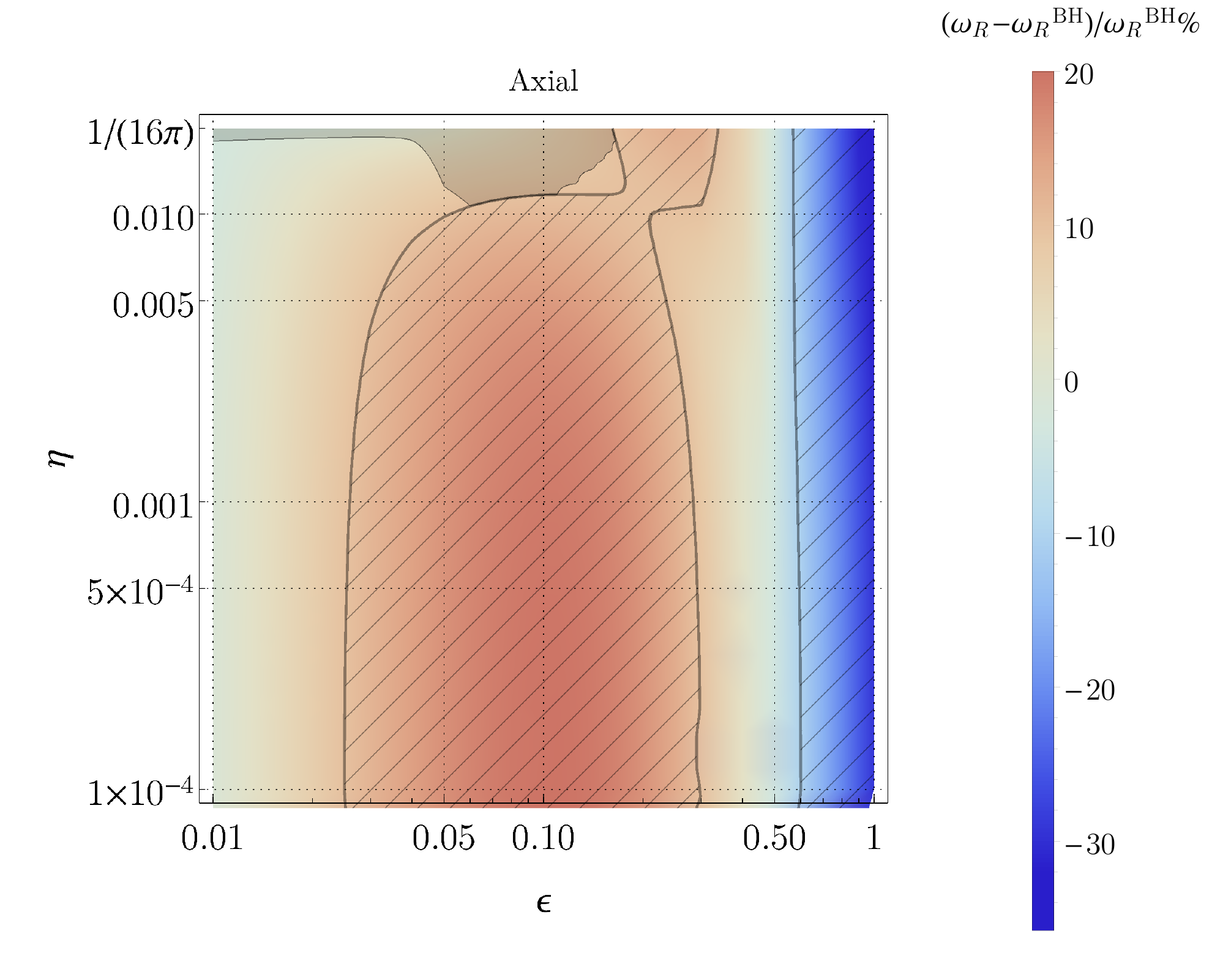}
\includegraphics[width=0.48\textwidth]{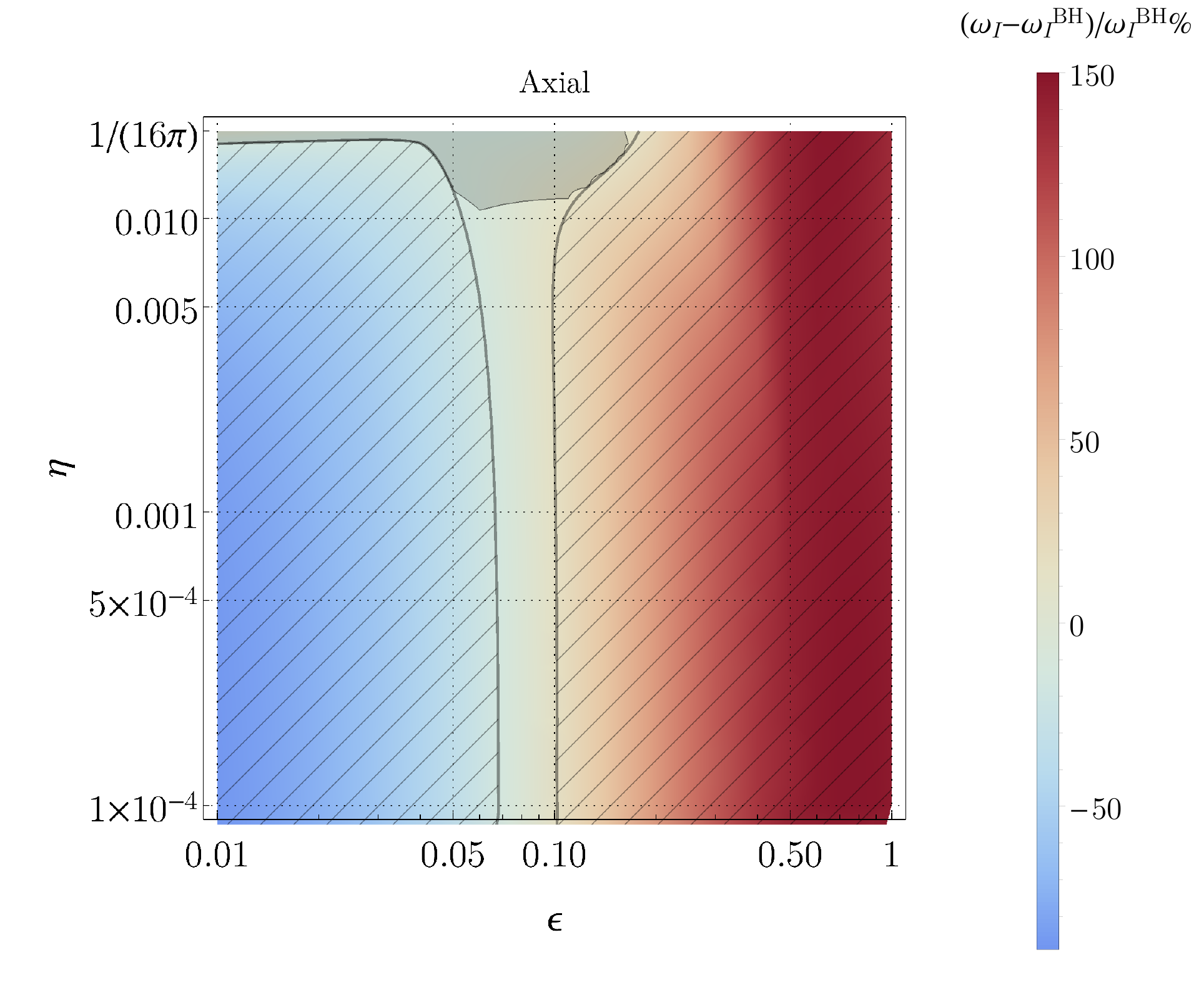}
\includegraphics[width=0.48\textwidth]{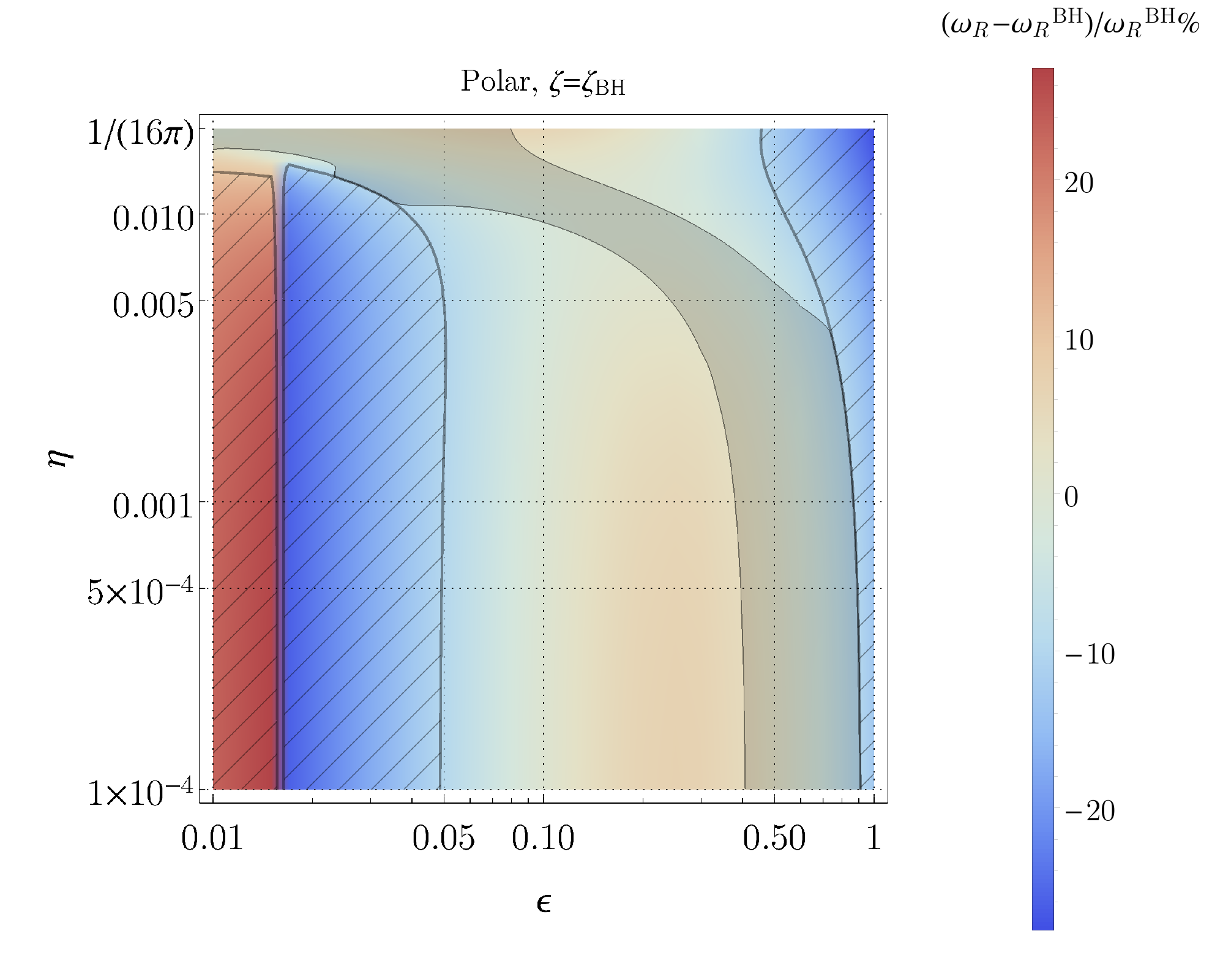}
\includegraphics[width=0.48\textwidth]{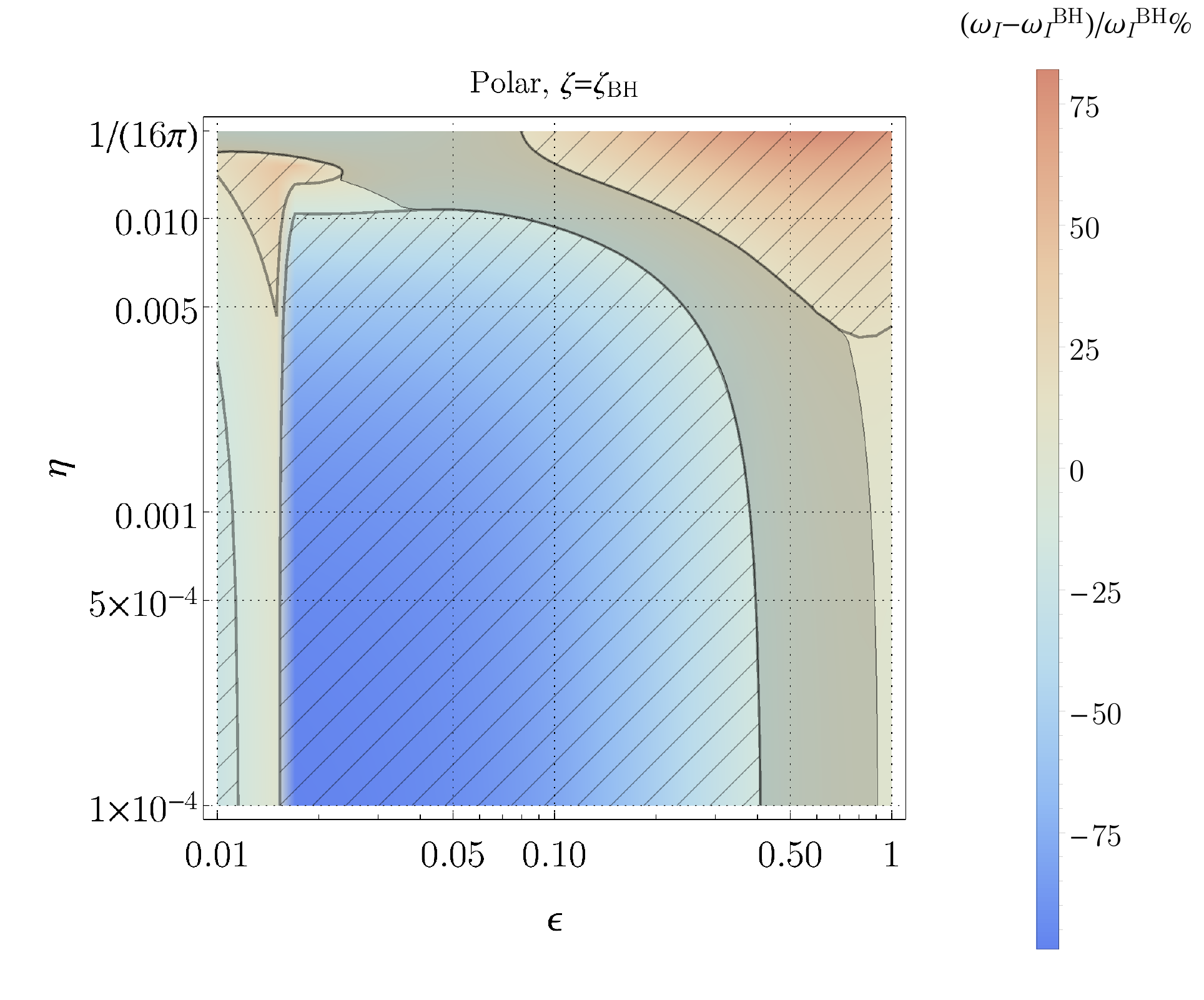}
\caption{Relative percentage difference of the real (left panel) and imaginary (right panel) part of the gravitational 
$l=2$ QNMs of a dark compact object with respect to the BH QNM. The difference is parametrized by the shear 
viscosity $\eta\in [0,\eta_{\rm BH}]$ of a fictitious fluid located at $R=2M(1+\epsilon)$ and by the compactness of the object through the 
parameter $\epsilon\in [0.01,1]$. As a reference, the contour lines correspond to the accuracy in the measurement of the
fundamental QNM of GW150914 and the dashed areas are the regions of the parameter space potentially excluded by 
individual measurements of $\omega_R$ (left) or $\omega_I$ (right). The dark shaded region in all panels is the area 
not excluded by a simultaneous measurement of the frequency and the damping time with the same accuracy as GW150914.
}
\label{fig:2dplot}
\end{figure*}

Figure~\ref{fig:2dplot} shows the relative percentage difference between the BH QNM and the QNMs of a dark compact object in the $(\epsilon,\eta$) parameter space. Based on the previous discussion, we 
focus on the region which directly affects the prompt ringdown, i.e., $\epsilon\gtrsim{\cal O}(0.01)$ and 
$\eta\in[0,\eta_{\rm BH}]$.

As a reference, in Fig.~\ref{fig:2dplot} we show the contour lines corresponding to $10\%$ and $15\%$ of relative 
difference for $\omega_R$ and $\omega_I$, respectively, which roughly correspond to the accuracy 
within which the fundamental QNM of the remnant in GW150914 has been measured. Although in the latter case the remnant is 
spinning~\cite{Abbott:2016blz}, we can assume $\approx(10-15)\%$ as an approximate 
estimate for the ringdown resolution of LIGO/Virgo for ``golden'' events such as GW150914.
For the same type of events, an overall improvement of the signal-to-noise ratio by an order of magnitude will provide 
a resolution at the percent level.

Measuring a single mode in the ringdown only allows to extract the mass and the spin of the object and one would need 
to measure at least two QNMs to verify whether the QNM spectrum is compatible with a BH in General Relativity. However, 
independent estimates of the mass and spin can be extracted from the inspiral-merger signal, assuming a binary BH in 
General Relativity~\cite{Abbott:2016blz,TheLIGOScientific:2016src}. Any deviation from the expected 
fundamental QNM would imply a failure of this test, suggesting either a departure from General Relativity or that the 
remnant is not a Kerr BH, or both.

Interestingly, already with the current LIGO/Virgo accuracy one can potentially place very strong constraints on the 
parameter space of our model, excluding the regions of the parameter space hatched in Fig.~\ref{fig:2dplot}. 
By combining the information from the real and the imaginary part of the QNMs, Fig.~\ref{fig:2dplot} shows that only a 
small region of the parameter space with $\epsilon \lesssim 0.1$ and $\eta \sim \eta_{\rm BH}$ (dark shaded area) is 
compatible with the current constraints for both axial and polar perturbations. For polar perturbations, also a region 
with $\epsilon \gtrsim 0.1$ and $0<\eta<\eta_{\rm BH}$ is compatible with the current constraints. 
Our results suggest that an improvement by a factor of a few in the detector sensitivity relative to LIGO first 
observational run would firmly rule out any modification of the prompt ringdown in our model.

\begin{figure}[t]
\centering
\includegraphics[width=0.48\textwidth]{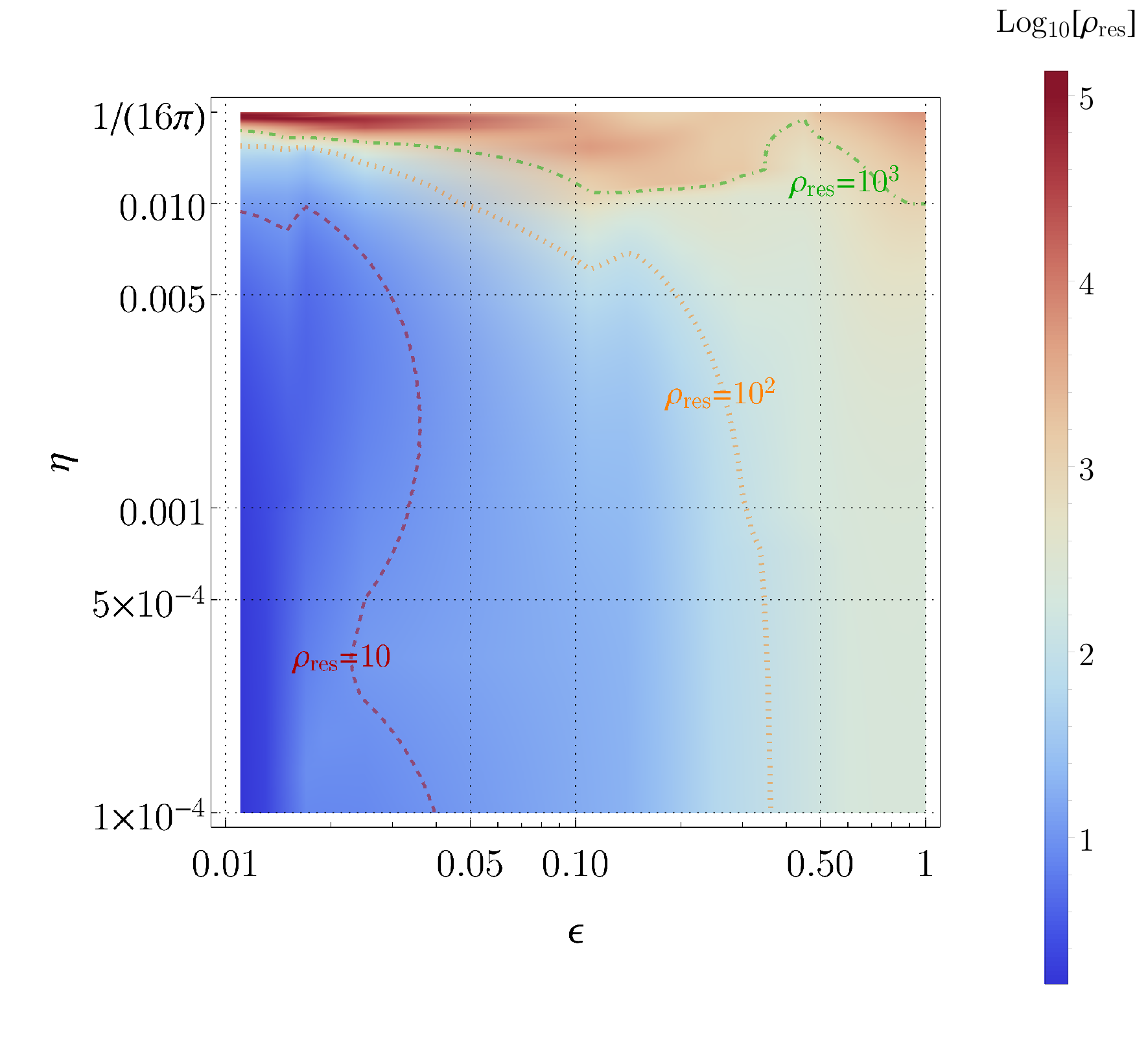}
\caption{Minimum signal-to-noise ratio required for the resolvability of the axial-polar QNM doublet according to the 
Rayleigh criterion.}
\label{fig:resolvability}
\end{figure}

Furthermore, as previously discussed, a generic feature of departure from the BH picture is the presence of a mode 
doublet. The frequencies of axial and polar modes are similar in most of the parameter space, with 
some regions featuring relative differences as large as $\approx 40\%$; however, the relative difference of the damping 
times can be much bigger (as large as $\approx 1500\%$ in some regions). 
A necessary condition for resolving these differences is based on the Rayleigh resolvability 
criterion~\cite{Berti:2005ys,Berti:2007zu}
\begin{eqnarray}
   \max[\sigma_{f_{1}}, \sigma_{f_{2}}]&<&|f_{1} - f_{2}|\,, \label{eq:resf}\\          
   \max[\sigma_{Q_{1}}, \sigma_{Q_{2}}]&<&|Q_{1} - Q_{2}|\,, \label{eq:restau} 
\end{eqnarray}
where $f_i=\omega_R^{(i)}/(2\pi)$ and $Q_i=\pi f_i \tau_i$ are the frequency and quality factor of the $i$-th mode, and 
$\sigma_{X}$ is the uncertainty in recovering 
a quantity $X$. 
In Fig.~\ref{fig:resolvability} we show the minimum signal-to-noise ratio, $\rho_{\rm res}$, required to resolve the 
axial and polar QNMs according to the above criterion. We computed the uncertainties on the parameters with a simple 
Fisher analysis~\cite{Berti:2005ys,Berti:2007zu} adapted from the case of BH 
overtones~\cite{Bhagwat:2019dtm,Forteza:2020hbw}, and assuming that the amplitude ratio between the axial and polar modes 
is $1/10$. In the $\eta\approx\eta_{\rm BH}$ region $\rho_{\rm res}>10^3$, but it 
can be much smaller for $\eta<\eta_{\rm BH}$. 
Although, a comparison between Fig.~\ref{fig:resolvability} and Fig.~\ref{fig:2dplot} shows that 
detecting deviations from the BH QNMs should require a much smaller signal-to-noise ratio than resolving the 
axial-polar doublet, the existence of this doublet is a generic smoking gun of objects other than BHs, so it 
might be useful to place model-independent constraints.

Finally, let us look at the opposite regime, where $\epsilon\ll0.01$. In this case the prompt ringdown is not modified 
and the constraints on the model rely on the detectability of the echoes. This problem has been recently analyzed in 
detail~\cite{Nakano:2017fvh,Mark:2017dnq,Bueno:2017hyj,Wang:2018mlp,Wang:2018gin,Uchikata:2019frs,Testa:2018bzd,
Maggio:2019zyv,Conklin:2019fcs,Tsang:2019zra}. In particular, Refs.~\cite{Testa:2018bzd,Maggio:2019zyv} analyzed the detectability as 
a function of the intrinsic reflectivity of the object, which coincides with Eq.~\eqref{R2} at high frequencies. By 
using the latter equation, we can map previous constraints on the object's reflectivity to the viscosity parameter.
In particular, using the results of Refs.~\cite{Testa:2018bzd,Maggio:2019zyv}, our analysis suggests that echoes are 
detectable with current GW interferometers at $2\sigma$ confidence level if $\eta<0.06 \eta_{\rm BH}$. Future detectors 
such as the Einstein Telescope~\cite{Hild:2010id} and LISA~\cite{Audley:2017drz} will allow to probe $\eta< 0.4\eta_{\rm 
BH}$ and $\eta< 0.6\eta_{\rm BH}$, respectively, at the same confidence level, while future LISA+ 
concepts~\cite{Baibhav:2019rsa} can reach almost $\eta\approx\eta_{\rm BH}$ at $3\sigma$ level. In other words, current 
GW interferometers could detect echoes only if the effective viscosity of the object is at most $6\%$ of that of a BH, 
larger values of $\eta$ would result in echoes that are too damped to be detectable with current facilities.

%%%%%%%%%%%%%%%%%%%%%%%%%%%%%%%%%%%%%%%%%%
\section{Concluding remarks}
%%%%%%%%%%%%%%%%%%%%%%%%%%%%%%%%%%%%%%%%%%

If an object is almost as compact as a BH and absorbs radiation almost as much as BH, then its QNM spectrum is almost 
indistinguishable from that of a BH. Besides confirming this reasonable expectation, our analysis provides a general 
framework to quantify the next natural question: 
\emph{when does the ringdown of a dark compact object become distinguishable from that of a BH?}

If an object with radius $R=2M(1+\epsilon)$ is sufficiently compact ($\epsilon\ll0.01$), the prompt ringdown is 
universal and the properties of the object's interior can appear only after the scrambling time 
$\tau_{\rm echo}\sim M|\log\epsilon|$~\cite{Cardoso:2016rao,Cardoso:2016oxy,Cardoso:2017cqb}, provided that the 
viscosity is sufficiently small. Thus, a negative echo search places an upper bound on the viscosity of the model, with 
current (future) detectors being able to probe values of $\eta$ as large as $6\%$ ($60\%$) of the BH viscosity.

For less compact objects ($\epsilon\gtrsim0.01$), there are no repeated echoes but the prompt ringdown is affected by 
the properties of the object's interior. In particular, the late-time ringdown is 
governed by the modified QNMs of the system. When the object is a perfect absorber at high frequency ($\eta=\eta_{\rm 
BH}$) the fundamental QNM deviates as much as ${\cal O}(10\%)$ and ${\cal O}(100\%)$ (for the real and imaginary part, 
respectively) from the BH case when $\epsilon={\cal O}(0.1)$. Although a more refined analysis is needed (in particular 
including the spin of the object), this suggests that inspiral-merger-ringdown consistency 
tests~\cite{TheLIGOScientific:2016src,LIGOScientific:2019fpa} can already put a strong constraint on $\epsilon$ of the 
order of $\epsilon\lesssim0.01$, i.e., the compactness of the remnant cannot be smaller than $99\%$ that of a 
BH (see Fig.~\ref{fig:2dplot}).
Future ringdown detections have the capability to rule out any deviation of the prompt ringdown in our model.

A generic feature that emerges from our framework is the fact that objects with $\epsilon\gtrsim{\cal O}(0.1)$ 
have a significant reflectivity at the typical ringdown frequency (even when $\eta=\eta_{\rm BH}$) compared to the BH 
case. This enhances the effects of the object's interior in the ringdown.
This phenomenon is most dramatic in the axial case, when the QNM spectrum becomes independent of $\eta$ as 
$\epsilon\to1/2$ and the reflectivity is unity. An object with radius equal to the light ring becomes a \emph{perfect 
reflector} of axial GWs and its axial QNM spectrum is universal.

Another generic feature is the breaking of isospectrality. Axial and polar modes are generically different except for 
the measure-one subspace ($\epsilon\to0$, $\eta=\eta_{\rm BH}$) of the whole three-dimensional $(\epsilon, \eta, 
\zeta)$ parameter space. Assuming that axial and polar modes are excited with comparable amplitude during the merger, 
this 
implies that a dark object other than a BH should vibrate with a mode \emph{doublet}. Resolving these doublet when 
$\eta\approx\eta_{\rm BH}$ requires a signal-to-noise ratio in the ringdown signal above $10^3$, whereas it takes much 
smaller sensitivity in other regions of the parameter space.

Our framework can be applied to specific models of extreme compact objects~\cite{Cardoso:2019rvt}, such as 
wormholes~\cite{Damour:2007ap}, gravastars~\cite{Mazur:2004fk,Visser:2003ge,Pani:2010em,Chirenti:2007mk}, 
superspinars~\cite{Gimon:2007ur}, nonlocal stars~\cite{Buoninfante:2019swn}, and horizonless microstate 
geometries~\cite{Myers:1997qi,Mathur:2005zp,Bena:2007kg,Balasubramanian:2008da,Bena:2013dka}.
In specific models the viscosity parameters might be complex and frequency dependent. This extension should be 
straightforward and is left for the future.

Likewise, we focused on nonrotating objects for simplicity, but our approach is conceptually valid also in the case of 
rotation. In that case, we envisage two technical obstacles: (i) the junction conditions require the analysis of 
metric perturbations, which should therefore be reconstructed from the Teukolsky 
function~\cite{Lousto:2002em,Whiting:2005hr}; 
(ii) the absence of Birkhoff's theorem in axisymmetry implies that the exterior of the object is not necessarily 
described by the Kerr metric. While deviations might be small in the $\epsilon\to0$ limit~\cite{Raposo:2018xkf}, 
when $\epsilon={\cal O}(0.1-1)$ a case-by-case analysis might be required. In the spinning case, it would also be 
interesting to study the ergoregion instability~\cite{Cardoso:2007az,Cardoso:2008kj,Pani:2010jz} and how the latter can 
be quenched in particular models~\cite{Maggio:2017ivp,Maggio:2018ivz,Oshita:2019sat}.

%%%%%%%%%%%%%%%%%%%%%%%%%%
\acknowledgments
The authors are grateful to Emanuele Berti for useful correspondence on the continued fraction method for stars, and
to Samir Mathur, Gabriel Andres Piovano, Guilherme Raposo, and Lucas Tonetto for useful discussions.
L.B. thanks Sapienza University of Rome for the hospitality in the first stages of this project.
L.B. acknowledges financial support from JSPS and KAKENHI Grant-in-Aid for Scientific Research No.~JP19F19164.
P.P. and E.M. acknowledge financial support provided under the European Union's H2020 ERC, Starting 
Grant agreement no.~DarkGRA--757480, under the MIUR PRIN and FARE programmes (GW-NEXT, CUP:~B84I20000100001), and 
support from the Amaldi Research Center funded by the MIUR program ``Dipartimento di 
Eccellenza" (CUP:~B81I18001170001).
A.M. is supported by Netherlands Organization for Scientific Research (NWO) Grant number 680-91-119.

\appendix

%%%%%%%%%%%%%%%%%%%%%%%%%%%%%%%%%%%%%%%%%%%%%%%%%%%
\section{Membrane paradigm}\label{app:paradigm}
%%%%%%%%%%%%%%%%%%%%%%%%%%%%%%%%%%%%%%%%%%%%%%%%%%%
In this appendix we provide details on the computation leading to the boundary conditions~\eqref{BC-axial} and 
\eqref{BC-polar}.
Let $\mathcal{M}$ be the whole spacetime manifold. We can define a $3$-dimensional spherical membrane (or shell) 
generated by time-like geodesics of radius $R=2M(1+\epsilon)$,
whose interior $\mathcal{M}^{-}$ and exterior $\mathcal{M}^{+}$ regions are described by two different metrics 
$g_{\mu\nu}^{-}(x^{-})$ and $g_{\mu\nu}^{+}(x^{+})$, respectively, with $x^{\pm}$ being the respective coordinates. The 
exterior region is described by the Schwarzschild metric~\eqref{backg-metric}.
The two sets of coordinates are
\begin{equation}
x^{-\mu}=(t^{-},r^{-},\theta^{-},\varphi^{-})\,,\qquad x^{+\mu}=(t^{+},r^{+},\theta^{+},\varphi^{+})\,,
\label{coordinates in-out}
\end{equation}
while the membrane's coordinates are $x_m^{\mu}=(t,R,\theta,\varphi)$, so that the intrinsic $3$-dimensional 
coordinates on the shell are 
\begin{equation}
y^{a}=(t,\theta,\varphi)\,.
\end{equation}
The normal vector $n^{\mu}$ to the time-like membrane is space-like, i.e., $n_{\mu}n^{\mu}=1$. A basis of three 
independent generators for the shell can be chosen as
\begin{equation}
e_a^{\mu}=\frac{\partial x^{\mu}}{\partial y^a}\,,
\label{basis}
\end{equation}
through which we can define the induced metric
\begin{equation}
h_{ab}=e^{\mu}_ae^{\nu}_bg_{\mu\nu}\,.
\label{induced metric}
\end{equation}
on the $3$-dimensional surface. The extrinsic curvature on the membrane is defined as
\begin{equation}
K_{ab}=e^{\mu}_ae^{\nu}_b\nabla_{\mu}n_{\nu}\,,
\label{extrins-curv}
\end{equation}
and its trace is $K=h^{ab}K_{ab}$.
To embed the $3$-dimensional membrane in the manifold $\mathcal{M}$ we need to impose a junction 
condition that glues the interior region with the exterior~\cite{Israel:1966rt}:
\begin{equation}
(K^{+}h_{ab}-K^{+}_{ab})-\left(K^{-}h_{ab}-K^{-}_{ab}\right)=8\pi T_{ab}\,,\label{israel1}
\end{equation}
where $K_{ab}^{+}$ and $K_{ab}^{-}$ are the extrinsic curvatures defined on the top and on the bottom of the shell, 
respectively; $T_{ab}$ is the stress-energy tensor of some matter distribution located on the membrane creating a 
discontinuity in the extrinsic curvature. 

The crucial and fundamental assumption of the membrane paradigm is that such a stress-energy tensor is fictitious and 
its role is to carry all the information about the interior. In the standard paradigm, one assumes that the inside 
region is such that
\begin{equation}
K_{ab}^{-}=0\,,\label{assumption}
\end{equation}
so that the junction condition~\eqref{israel1} reduces to~\cite{Thorne:1986iy}
\begin{equation}
Kh_{ab}-K_{ab}=8\pi T_{ab}\,,\label{israel-thorne}
\end{equation}
where for simplicity we have redefined $K_{ab}\equiv K_{ab}^+,$ and henceforth we denote the coordinates outside simply as $x^{\mu}=(t,r,\theta,\varphi)$. 

The junction conditions are compatible with the stress-energy tensor of a dissipative 
fluid, Eq.~\eqref{stress-energy},
where $\rho$ and $p$ are the density and pressure, respectively; $u_a$ is the $3$-velocity of the fluid defined in 
terms of its $4$-velocity $U_{\mu}$ as $u_a=e^{\mu}_aU_{\mu}$. Moreover, $\Theta=u^a_{;a}$ is the expansion, 
$\sigma_{ab}=\frac{1}{2}\left(u_{a;c}\gamma^{c}_b+u_{b;c}\gamma^{c}_a-\Theta \gamma_{ab}\right)$
is the shear tensor, where $\gamma_{ab}=h_{ab}+u_au_b$ is the projector tensor, and $u_{b;a}$ is the $3$-dimensional 
covariant derivative compatible with the induced metric $h_{ab}$~\cite{Kojima:1992ie,Uchikata:2016qku}.

\subsection{Unperturbed background}

In the static background [Eq.~\eqref{backg-metric}] the viscosity parameters $\eta$ and $\zeta$ do not play any 
role. Indeed, it is easy to show that the expansion and shear are zero, so the membrane stress-energy tensor reduces to 
that a perfect fluid. In this case, the $3$-metric reads
\begin{equation}
h_{tt}=-f(R)\,,\quad h_{\theta\theta}=\frac{h_{\varphi\varphi}}{{\rm sin}^2\theta}=R^2\,.
\label{3-metric backg}
\end{equation}
The normal vector $n^\mu$ can be uniquely determined by imposing the four conditions $n^\mu n_\mu=1$ and 
$e_{a}^\mu n_\mu=0$; its components read
\begin{equation}
n_{t}=0\,,\quad n_r=\frac{1}{\sqrt{f(r)}}\,,\quad n_{\theta}=0,\quad n_{\varphi}=0.\label{normal vector-down}
\end{equation}
The extrinsic curvature is also diagonal,
\begin{eqnarray}
K_{tt}= -\frac{1}{2} \sqrt{f(R)} f^{\prime}(R)\,,\quad 
K_{\theta\theta}=\frac {K_{\varphi\varphi} }{\sin^2 \theta}= R\sqrt{f(R)}\,, \nonumber
\end{eqnarray}
where a prime denotes partial derivative with respect to the argument.

The only nonvanishing component of the fluid $4$-velocity 
$U^\mu=(U^t,U^{r},U^\theta,U^\varphi)$ is $U^t=1/\sqrt{f(R)}$. This yields the
the $3$-velocity  
\begin{equation}
u_a=e^{\mu}_a U_{\mu}\,,\qquad u^a=h^{ab}u_b= \left(\frac{1}{\sqrt{f(R)}},0,0\right)\,.
\end{equation}
Therefore, the nonvanishing components of the junction condition~\eqref{israel-thorne} at the background level are
\begin{eqnarray}
tt: \,\,&&  -\frac{2}{R}f^{3/2}(R)=8\pi f(R) \rho_0 \,, \nonumber\\
\theta\theta:\,\, && \frac{R\left(2f(R)+Rf^{\prime}(R)\right)}{2\sqrt{f(R)}}=8\pi R^2p_0\,, \nonumber\\
\varphi\varphi:\,\, && \frac{R{\rm sin}^2\theta\left(2f(R)+rf^{\prime}(R)\right)}{2\sqrt{f(R)}}=8\pi R^2{\rm 
sin}^2\theta p_0\,.\qquad
\end{eqnarray}
The $(tt)$ component gives the unperturbed density of the membrane~\cite{Abedi:2020ujo}
\begin{equation}
\rho_0(R)=-\frac{\sqrt{f(R)}}{4\pi R}\,;
\label{rho0}
\end{equation}
while the angular components give the unperturbed pressure of the 
membrane
\begin{equation}
p_0(R)=\frac{2f(R)+Rf^{\prime}(R)}{16\pi R\sqrt{f(R)}}\,.
\label{p0}
\end{equation}
The speed of sound is 
%%%
\begin{equation}
	c_s \equiv \sqrt{\frac{\partial p_0}{\partial \rho_0}} = \sqrt{\frac{1+2 \epsilon +4 \epsilon ^2}{8 \epsilon  
(1/2-\epsilon )}}\,,
	\label{sound-speed}
\end{equation}
%%%
which diverges as $\epsilon\to0,1/2$ and is complex for $\epsilon>1/2$. Note, however, that the properties of the 
fluid do not need to be physical, since the membrane is fictitious.

\subsection{Gravitational perturbations}

We work in the Regge-Wheeler gauge~\cite{Regge:1957td} and study separately the axial and polar 
sectors of gravitational perturbations, since they decouple in presence of a static background.

The perturbed metric can be cast in the following general form
\begin{equation}
g_{\mu\nu}=g_{\mu\nu}^{0}(r)+\delta g_{\mu\nu}(r,\theta,t)\,,
\end{equation}
where, without loss of generality, the perturbation $\delta 
g_{\mu\nu}$ does not depend on the azimuthal angle $\varphi$, owing to the spherical symmetry of the background 
$g_{\mu\nu}^{0}$.

Because of the metric perturbations, the dissipative components of the stress-energy tensor are switched on, and both 
density and pressure are perturbed as follows
\begin{eqnarray}
\rho&=&\rho_0+\delta \rho(t,\theta)\,,\\
p&=&p_0+\delta p(t,\theta)\,.
\end{eqnarray}
Also the location of the membrane is affected by the perturbation, and we can parametrize the 
deviation as
\begin{equation}
r_m(t,\theta)=R+\delta R(t,\theta)\,.
\end{equation}

We can also easily find the expressions of the perturbed tangential vectors $e^{\mu}_a$ introduced in Eq.~\eqref{basis}. 
Since the $4$-dimensional coordinates of the membrane are
\begin{equation}
x_m^{\mu}=\left(t,R+\delta R(t,\theta),\theta,\varphi\right)\,,
\end{equation}
we have
\begin{eqnarray}
e^{\mu}_t&=&\left(1,\partial_t \delta R(t,\theta),0,0\right)\,,\nonumber\\
e^{\mu}_\theta&=&\left(0,
\partial_\theta \delta R(t,\theta),1,0\right)\,,\nonumber\\
e^{\mu}_\varphi&=&\left(0,0,0,1\right)\,.
\label{perturbed-basis}
\end{eqnarray}

Note that, $\delta \rho(t,\theta)$, $\delta p(t,\theta)$ and $\delta R(t,\theta)$ are scalar quantities under 
rotation, so they are only affected by polar perturbations and can be decomposed as
\begin{eqnarray}
\delta \rho (t,\theta)&=&\varepsilon\rho_1 P_l(\cos\theta) e^{-i\omega t}\,,\nonumber\\
\delta p(t,\theta)&=&\varepsilon p_1 P_l(\cos\theta) e^{-i\omega t}\,,\nonumber\\
%%%
\delta R(t,\theta)&=&\varepsilon \delta R_0  P_l(\cos\theta)  e^{-i\omega t}\,. \label{deltaR}
\end{eqnarray}
where $\rho_1$, $p_1$, $\delta R_0$ only depend on the unperturbed radius $R$, $P_l(\cos \theta)$ are the Legendre 
polynomials\footnote{Note that we can set $m=0$ in the spherical harmonics $Y_l^m(\theta,\varphi)\propto 
e^{im\varphi}P_l^m(\cos \theta)$ without any loss of generality due to the spherical symmetry of the background.} and 
the parameter $\varepsilon$ keeps track of 
the order in perturbation, so that all contributions of order 
$\mathcal{O}(\varepsilon^2)$ are negligible.

In the following we treat the axial and polar sectors separately.

\subsubsection{Axial sector}

The nonvanishing components of the axial metric perturbations in the Regge-Wheeler gauge~\cite{Regge:1957td} are
\begin{eqnarray}
\delta g_{t\varphi}&=&\varepsilon e^{-i\omega t}h_0(r)\sin \theta  \partial_\theta P_l(\cos \theta)\,,\nonumber\\
\delta g_{r\varphi}&=&\varepsilon e^{-i\omega t}h_1(r)\sin \theta  \partial_\theta P_l(\cos \theta)\,.
\label{non-zero component}
\end{eqnarray}
It follows that the only nonvanishing component of the $3$-metric perturbation $\delta h_{ab}$ is
\begin{equation}
 \delta h_{t\varphi}=\varepsilon e^{-i\omega t}h_0(R)\sin \theta  \partial_\theta P_l(\cos \theta)\,.
\label{3-metric down}
\end{equation}
In the case of axial perturbations the the four components of the normal vector up to ${\cal O}(\varepsilon)$ are 
simply given by the unperturbed one in Eq.~\eqref{normal vector-down}. As a consequence, the nonvanishing components 
of the extrinsic curvature perturbation are 
\begin{eqnarray}
\delta K_{t\varphi}&=&  \frac{1}{2}e^{-i\omega t}\varepsilon \sqrt{f}\left(i\omega 
h_1+h_0^{\prime}\right)\sin\theta \partial_\theta P_l(\cos\theta)\,, \nonumber \\
\delta K_{\theta\varphi}&=&-\frac{1}{2}e^{-i\omega t}\varepsilon \sqrt{f}h_1\left(\cos\theta \partial_\theta 
+\sin\theta \partial^2_\theta \right)P_l(\cos\theta)\,.\nonumber
\end{eqnarray}

As for the fluid velocity, the components $U^t$, $U^r$ and $U^\theta$ are not affected by axial perturbations, while 
$\delta u^\varphi\neq 0$ and its expression can be found by solving the $(t\varphi)$ component of the junction 
condition,
\begin{eqnarray}
\delta u^{\varphi}= \frac{\varepsilon e^{-i\omega t}\partial_\theta 
P_l(\cos \theta)\left[h_0f^{\prime}-f\left(i\omega h_1+ h_0^{\prime}\right)\right] 
}{R{\rm sin}\theta \sqrt{f}\left(2f-Rf^{\prime} \right)} \,.\qquad
\end{eqnarray}
The perturbed components of the stress-energy tensor are
\begin{eqnarray}
\delta T_{t\varphi}&=& - \varepsilon e^{-i \omega t}\rho_0 h_0 \sin\theta\partial_\theta P_l(\cos\theta )\nonumber\\
&&\nonumber+R^2 \sqrt{f} \sin^2\theta (p_0+\rho_0) \delta u^\varphi(t,\theta )\,,\\
\delta T_{\theta\varphi}&=& -\eta \, R^2 \sin^2\theta  \partial_\theta \delta u^\varphi(t,\theta ) \,.
\end{eqnarray}
The $(\theta\varphi)$ component of the junction condition then reduces to
\begin{eqnarray}
\frac{1}{2}\sqrt{f}h_1=\frac{8\pi\eta  R\left[h_0f^{\prime}-f\left(i\omega 
h_1+ h_0^{\prime}\right)\right]}{\sqrt{f}\left(2f-Rf^{\prime} \right)}\,.
\label{theta-phi}
\end{eqnarray}
In vacuum the Regge-Wheeler functions are related to each other~\cite{Regge:1957td}, 
$h_0(r)=-\frac{f(r)}{i\omega}\frac{{\rm d}}{{\rm d}r}\left[f(r)h_1(r)\right]$. One can use this relation to eliminate 
$h_0(R)$ and $h_0^{\prime}(R)$ in terms of $h_1(R)$, $h_1^{\prime}(R)$ and 
$h_1^{\prime\prime}(R)$.
Furthermore, one can replace $h_1$ and its derivatives by introducing the Regge-Wheeler 
function,
\begin{equation}
\psi(r)=\frac{f(r)}{r}h_1(r) \label{psih1}
\end{equation}
which satisfies Eq.~\eqref{ODE} with effective potential $V_{\rm axial}$. We finally obtain
\begin{equation}
\omega \psi(R) = i16 \pi \eta  \left( \left.\frac{\partial \psi}{\partial x}\right|_{R} + 
\frac{RV_{\rm axial}(R)}{2f(R)-Rf^{\prime}(R)} \psi(R) \right) \,.
\label{BC-eta-r02}
\end{equation}
which coincides with Eq.~\eqref{BC-axial} in the main text.

\subsubsection{Polar sector}

The nonvanishing components of the polar metric perturbation are
\begin{eqnarray}
\delta g_{tt}&=&\nonumber\varepsilon e^{-i\omega t}P_l(\cos\theta)f(r){\cal H}(r)\,,\\
\delta g_{rr}&=&\nonumber\varepsilon e^{-i\omega 
t}P_l(\cos\theta)\frac{{\cal H}(r)}{f(r)}\,, \\
\delta g_{tr}&=&\nonumber\varepsilon e^{-i\omega t}P_l(\cos\theta){\cal H}_1(r)\,,\\
\delta g_{\theta\theta}&=&\frac{\delta 
g_{\varphi\varphi}}{{\rm sin}^2\theta}=\varepsilon e^{-i\omega t}P_l(\cos\theta)r^2{\cal K}(r)\,,
\label{non-zero component-polar}
\end{eqnarray}
and the location of the membrane is perturbed as in Eq.~\eqref{deltaR}.

By projecting on the $3$-dimensional membrane, the nonvanishing components of the $3$-metric 
perturbation $\delta h_{ab}$ read 
\begin{eqnarray}
\delta h_{tt}&=&\varepsilon \left(f {\cal H}-f^{\prime}\delta R_0\right) P_l(\cos\theta) e^{-i\omega t}\,,\nonumber\\
\delta h_{\theta\theta}&=&\frac{h_{\varphi\varphi}}{{\rm sin}^2\theta}=\varepsilon \left(R^2{\cal K}+2R \delta R_0\right)  
P_l(\cos\theta) e^{-i\omega t}\,.\qquad\quad  
\label{3-metric down-polar}
\end{eqnarray}
Moreover, the perturbed components of the normal vector for polar perturbations up to first order in $\varepsilon$ are
\begin{eqnarray}
\delta n_{t}&=&\frac{\varepsilon\,i\omega e^{-i\omega t}P_l(\cos\theta) \delta R_0}{\sqrt{f(r)}}\,,\nonumber\\
\delta 
n_r&=&\frac{\varepsilon e^{-i\omega t}P_l(\cos\theta){\cal H}(r)}{2\sqrt{f(r)}}\,,\nonumber\\
\delta n_{\theta}&=&-\frac{\varepsilon 
e^{-i\omega t} \partial_\theta P_l(\cos\theta)\delta R_0}{\sqrt{f(r)}}\,.\label{normal vector-down-polar}
\end{eqnarray}
As for the perturbed fluid velocity, in the polar sector $\delta U^t$ can be uniquely determined from the condition on the norm
$U_\mu U^\mu=-1,$
\begin{equation}
\delta U^t=\delta u^t=\frac{\varepsilon \left(f{\cal H}- \delta R_0f^{\prime}\right)}{2f^{3/2}}P_l(\cos\theta) e^{-i\omega 
t}\,.
\end{equation}
Moreover, $\delta U^{\varphi}=\delta u^\varphi=0$ and $\delta U^r=-\varepsilon U^t\,i\omega  e^{-i\omega 
t}P_l(\cos\theta)$, while $\delta U^\theta=\delta u^\theta$ is nonvanishing and can be determined by solving the 
$(t\theta)$ component of the junction condition~\eqref{israel-thorne}, as we will see below.  

\begin{widetext}
		
We can now compute the polar perturbation in the extrinsic curvature whose nonvanishing components up to first order 
in $\varepsilon$ are 
\begin{eqnarray}
\delta K_{tt}&=&\frac{\varepsilon  e^{-i \omega t}  P_l(\cos\theta ) 
\left[\delta R_0 \left(4 \omega^2-{f^{\prime}}^2\right)+f 
\left(-2 \delta R_0 f^{\prime\prime}+3 {\cal H} f^{\prime}+4 i \omega {\cal H}_1\right)+2 f^2 {\cal H}^{\prime}\right]}
{4\sqrt{f}}\,,\nonumber\\
\delta K_{\theta\theta}&=& \frac{\varepsilon e^{-i \omega t} \left[f \left( 2 \delta R_0-R {\cal H}+R^2 {\cal K}^\prime+2 R 
{\cal K}\right)+\delta R_0 \left(R  f^\prime-2 \partial_\theta^2\right)\right]P_l(\cos\theta )}{2 
\sqrt{f}}\,,\nonumber\\
\delta K_{\varphi\varphi}&=& \nonumber \frac{\varepsilon\sin^2\theta  e^{-i \omega t} \left[ f \left(2 \delta R_0-R {\cal H}+R^2 
{\cal K}^\prime+2 R {\cal K}\right)+\delta R_0 \left(R  f^\prime-2 \cot\theta  
\partial_\theta\right)\right]P_l(\cos\theta )}{2 \sqrt{f}} \,,\\
\delta K_{t\theta}&=&-\frac{\varepsilon  e^{-i \omega t} \partial_\theta P_l(\cos\theta ) (f {\cal H}_1-2 i \delta R_0 \omega 
)}{2 \sqrt{f}}\,.
\end{eqnarray}
The nonvanishing components of the perturbation to the stress-energy tensor are 
\begin{eqnarray}
\delta T_{tt}&=&  \varepsilon e^{-i\omega t} P_l(\cos\theta )  \left[\rho_0 f^{\prime} \delta R_0+ f \left(\rho 
_1-\rho_0 {\cal H}\right)\right]\,, \nonumber\\
\delta T_{\theta\theta}&=& \nonumber \frac{R }{\sqrt{f}} \left\lbrace -\sqrt{f}R  \left[  (\zeta +\eta ) 
\partial_\theta \delta u^\theta +  (\zeta -\eta ) \cot \theta  \delta u^\theta \right]\right. 
\nonumber\\
&&\left.+\varepsilon e^{-i  \omega t}P_l(\cos\theta) \left[\sqrt{f}\left(p_0 R {\cal K}+2 p_0 \delta R_0+p_1 R\right)+i  
\zeta  \omega  (R {\cal K}+2 \delta R_0)\right]\right\rbrace\,, \nonumber\\
\delta T_{\varphi\varphi}&=& \nonumber \frac{R\sin^2\theta }{\sqrt{f}} \left\lbrace -\sqrt{f}R  \left[  
(\zeta -\eta ) \partial_\theta \delta u^\theta +  (\zeta +\eta ) \cot \theta  \delta u^\theta \right]\right.\nonumber \\
&&\left.+\varepsilon e^{-i  \omega t}P_l(\cos\theta) \left[\sqrt{f}\left(p_0 R {\cal K}+2 p_0 \delta R_0+p_1 R\right)+i  
\zeta  \omega  (R {\cal K}+2 \delta R_0)\right]\right\rbrace\,,\nonumber\\
\delta T_{t\theta}&=& - R^2 \sqrt{f}(\rho_0+p_0)\delta u^\theta  \,.
\end{eqnarray}
\end{widetext}
From the components $(tt)$, $(\theta\theta)$, and $(\varphi\varphi)$ of the junction conditions we obtain 
analytical (albeit cumbersome) expressions for $\rho_1$ and $p_1,$ and the deviation in the membrane location
\begin{equation}
\delta R_0=\frac{16\pi \eta R f {\cal H}_1}{2f-Rf^\prime -32\pi \eta i\omega R} \,,
\end{equation}
whereas from the $(t\theta)$ component we derive
\begin{equation}
\delta u^\theta=\frac{\varepsilon e^{-i\omega t}\partial_\theta P_l(\cos \theta)R\sqrt{f}{\cal H}_1}{2f-Rf^\prime -32\pi \eta i\omega R}\,.
\end{equation}

This still leaves the metric perturbations, ${\cal H}(r)$, ${\cal H}_1(r)$ and ${\cal K}(r)$, unknown. In vacuum, these three functions are 
related by the following algebraic equation~\cite{Zerilli:1970se,Zerilli:1971wd}:
\begin{eqnarray}
{\cal H}(r)&=&\frac{1}{qr+3M}\left\lbrace \left[qr-\frac{\omega^2r^4}{r-2M}+M\frac{r-3M}{r-2M}\right]{\cal K}(r) \right.\nonumber\\
&&\left.+\left[ 
i\omega 
r^2+\frac{(q+1)M}{i\omega r} \right]{\cal H}_1(r)  \right\rbrace \,,
\label{algebraic eq.}
\end{eqnarray}
where $q=(l-1)(l+2)/2$. The last relation allows us to eliminate, say, ${\cal H}(r)$.

Moreover, we can rewrite ${\cal H}_1(r)$ and ${\cal K}(r)$ in terms of the Zerilli function $Z(x)$, which satisfies Eq.~\eqref{ODE} 
with $V=V_{\rm polar}$~\cite{Zerilli:1970se,Zerilli:1971wd}. Indeed,
\begin{eqnarray}
{\cal H}_1(r)&=&\omega h(r)Z(x)+\omega k(r)\frac{{\rm d}Z(x)}{{\rm d}x}\,,\\
{\cal K}(r)&=& g(r)Z(x)+\frac{{\rm d}Z(x)}{{\rm d}x}\,,
\label{rel-psi-polar}
\end{eqnarray}
%%%
where~\cite{Zerilli:1970se,Zerilli:1971wd}
\begin{eqnarray}
h(r) &=&  i\frac{3qMr-qr^2+3M^2}{(r-2M)(qr+3M)}\,, \nonumber\\
k(r) &=& -i\frac{r^2}{r-2M} \,,\nonumber \\
g(r)&=&\frac{q(q+1)r^2+3qMr+6M^2}{r^2(qr+3M)}\,.
\label{coeff-zerilli}
\end{eqnarray}

The last condition which closes the system of equations and uniquely determines the boundary conditions 
for the polar metric perturbations at $r=R$ is found from the barotropic equation of state $p=p(\rho)$, at first order 
in perturbation, which gives
\begin{equation}
\delta p=c_s^2\delta \rho \,,\label{barotrpic}
\end{equation}
where the sound speed squared is given by Eq.~\eqref{sound-speed}.

By substituting the above algebraic equations, after a tedious but conceptually simple calculation we obtain the 
following boundary condition
\begin{equation}
\frac{\partial_x Z(x)}{Z(x)}=-16\pi\eta i\omega+G(R,\omega,\eta,\zeta) \,,\label{BC-barotr}
\end{equation}
where $G={A}/{B}$ and
\begin{widetext}
\begin{eqnarray}
A&=&\nonumber (y-2) \left\lbrace 9 \left[-3-3 w^2 y^3+w^2 y^4-48 i \pi  w y^2 \zeta +y (2+96 i \pi  w \zeta 
)\right]+q^3 y^2\left(-3 i+i y+16 \pi  w y^2 \eta \right)^2\right.\\
&&\nonumber+3 q \left[-9+3 y^2 (1+64 i \pi  w\zeta )-32 i \pi  w y^3 (3 \zeta +2 \eta )+w y^4 (-3 w+16 i \pi  \eta 
)+w^2 
y^5 \left(1+256 \pi ^2 \eta^2\right)\right]\\
&&\left.+q^2 y \left[-18+6 y^2 (1+16 i \pi  w (\zeta -2 \eta ))+256 \pi ^2w^2 y^5 \eta ^2+32 \pi  w y^4 \eta  (i+24 \pi 
w \eta )-i y^3 (-i+48 \pi  w (\zeta +\eta ))\right]\right\rbrace 
\,,
\qquad\quad\\
B&=&\nonumber y^2 (3+qy) \left\lbrace 3 \left[-3-3 w^2 y^3+w^2 y^4-48 i \pi  w y^2 \zeta +y (2+96 i\pi  w \zeta 
)\right]\right.\\
&&\nonumber +q \left[-9+9 y+w^2 y^5+y^2 (-3+192 i \pi  w \zeta )+y^3 (1-32 i \pi  w (3 \zeta -\eta ))-w y^4(3 w+16 i 
\pi 
 \eta )\right]\\
&&\left.+q^2 y \left[3-3 y-16 i \pi  w y^3 (\zeta+\eta )+y^2 (1+32 i \pi  w (\zeta +\eta ))\right]\right\rbrace \,,
\end{eqnarray}
\end{widetext}
and we have defined the dimensionless quantities $y=R/M$, $w=M\omega$. Note that in the BH limit
$G(2M)=0$, and the BH boundary condition is recovered for $\eta=1/(16\pi)$, consistently with the axial case.

Finally, from our computations it follows that both the density $\rho$ and the 
expansion $\Theta$ vanish for a perturbed Schwarzschild spacetime.
In the BH limit the effective stress-energy tensor~\eqref{stress-energy}  reduces to
	\begin{equation}
	T_{ab}^{\rm BH}=p \gamma_{ab}-2\eta_{\rm BH} \sigma_{ab}\,,\qquad R\rightarrow 2M\,,
	\end{equation}
which is independent of $\zeta$, as discussed in the main text.

%----------------------------------------------------------------------------------------------------
\bibliographystyle{utphys}
\bibliography{References}

\providecommand{\href}[2]{#2}\begingroup\raggedright\begin{thebibliography}{100}

\bibitem{LambShift}
W.~E. Lamb and R.~C. Retherford, ``Fine structure of the hydrogen atom by a
  microwave method,'' \href{http://dx.doi.org/10.1103/PhysRev.72.241}{{\em
  Phys. Rev.} {\bfseries 72} (Aug, 1947) 241--243}.
  \url{https://link.aps.org/doi/10.1103/PhysRev.72.241}.

\bibitem{TheLIGOScientific:2016src}
{\bfseries LIGO Scientific, Virgo} Collaboration, B.~Abbott {\em et~al.},
  ``{Tests of general relativity with GW150914},''
  \href{http://dx.doi.org/10.1103/PhysRevLett.116.221101}{{\em Phys. Rev.
  Lett.} {\bfseries 116} no.~22, (2016) 221101},
  \href{http://arxiv.org/abs/1602.03841}{{\ttfamily arXiv:1602.03841 [gr-qc]}}.
  [Erratum: Phys.Rev.Lett. 121, 129902 (2018)].

\bibitem{LIGOScientific:2019fpa}
{\bfseries LIGO Scientific, Virgo} Collaboration, B.~Abbott {\em et~al.},
  ``{Tests of General Relativity with the Binary Black Hole Signals from the
  LIGO-Virgo Catalog GWTC-1},''
  \href{http://dx.doi.org/10.1103/PhysRevD.100.104036}{{\em Phys. Rev. D}
  {\bfseries 100} no.~10, (2019) 104036},
  \href{http://arxiv.org/abs/1903.04467}{{\ttfamily arXiv:1903.04467 [gr-qc]}}.

\bibitem{Vishveshwara:1970cc}
C.~V. Vishveshwara, ``{Stability of the schwarzschild metric},''
\href{http://dx.doi.org/10.1103/PhysRevD.1.2870}{{\em Phys. Rev.} {\bfseries
  D1} (1970) 2870--2879}.
%%CITATION = PHRVA,D1,2870;%%.

\bibitem{ChandraBook}
S.~Chandrasekhar, {\em {The mathematical theory of black holes}}.
\newblock 1985.

\bibitem{1980ApJ...239..292D}
S.~{Detweiler}, ``{Black holes and gravitational waves. III - The resonant
  frequencies of rotating holes},''
  \href{http://dx.doi.org/10.1086/158109}{{\em \apj} {\bfseries 239} (July,
  1980) 292--295}.

\bibitem{Dreyer:2003bv}
O.~Dreyer, B.~J. Kelly, B.~Krishnan, L.~S. Finn, D.~Garrison, and
  R.~Lopez-Aleman, ``{Black hole spectroscopy: Testing general relativity
  through gravitational wave observations},''
  \href{http://dx.doi.org/10.1088/0264-9381/21/4/003}{{\em Class. Quant. Grav.}
  {\bfseries 21} (2004) 787--804},
\href{http://arxiv.org/abs/gr-qc/0309007}{{\ttfamily arXiv:gr-qc/0309007
  [gr-qc]}}.
%%CITATION = GR-QC/0309007;%%.

\bibitem{Berti:2005ys}
E.~Berti, V.~Cardoso, and C.~M. Will, ``{On gravitational-wave spectroscopy of
  massive black holes with the space interferometer LISA},''
  \href{http://dx.doi.org/10.1103/PhysRevD.73.064030}{{\em Phys. Rev.}
  {\bfseries D73} (2006) 064030},
\href{http://arxiv.org/abs/gr-qc/0512160}{{\ttfamily arXiv:gr-qc/0512160
  [gr-qc]}}.
%%CITATION = GR-QC/0512160;%%.

\bibitem{Kokkotas:1999bd}
K.~D. Kokkotas and B.~G. Schmidt, ``{Quasinormal modes of stars and black
  holes},'' \href{http://dx.doi.org/10.12942/lrr-1999-2}{{\em Living Rev. Rel.}
  {\bfseries 2} (1999) 2},
\href{http://arxiv.org/abs/gr-qc/9909058}{{\ttfamily arXiv:gr-qc/9909058
  [gr-qc]}}.
%%CITATION = GR-QC/9909058;%%.

\bibitem{Berti:2009kk}
E.~Berti, V.~Cardoso, and A.~O. Starinets, ``{Quasinormal modes of black holes
  and black branes},''
  \href{http://dx.doi.org/10.1088/0264-9381/26/16/163001}{{\em Class. Quantum
  Grav.} {\bfseries 26} (2009) 163001},
\href{http://arxiv.org/abs/0905.2975}{{\ttfamily arXiv:0905.2975 [gr-qc]}}.
%%CITATION = ARXIV:0905.2975;%%.

\bibitem{Abbott:2016blz}
{\bfseries The LIGO/Virgo Scientific Collaboration} Collaboration, B.~P. Abbott
  {\em et~al.}, ``{Observation of Gravitational Waves from a Binary Black Hole
  Merger},'' \href{http://dx.doi.org/10.1103/PhysRevLett.116.061102}{{\em Phys.
  Rev. Lett.} {\bfseries 116} no.~6, (2016) 061102},
\href{http://arxiv.org/abs/1602.03837}{{\ttfamily arXiv:1602.03837 [gr-qc]}}.
%%CITATION = ARXIV:1602.03837;%%.

\bibitem{Berti:2015itd}
E.~Berti {\em et~al.}, ``{Testing General Relativity with Present and Future
  Astrophysical Observations},''
  \href{http://dx.doi.org/10.1088/0264-9381/32/24/243001}{{\em Class. Quant.
  Grav.} {\bfseries 32} (2015) 243001},
\href{http://arxiv.org/abs/1501.07274}{{\ttfamily arXiv:1501.07274 [gr-qc]}}.
%%CITATION = ARXIV:1501.07274;%%.

\bibitem{Barack:2018yly}
L.~Barack {\em et~al.}, ``{Black holes, gravitational waves and fundamental
  physics: a roadmap},''
\href{http://arxiv.org/abs/1806.05195}{{\ttfamily arXiv:1806.05195 [gr-qc]}}.
%%CITATION = ARXIV:1806.05195;%%.

\bibitem{Berti:2018vdi}
E.~Berti, K.~Yagi, H.~Yang, and N.~Yunes, ``{Extreme Gravity Tests with
  Gravitational Waves from Compact Binary Coalescences: (II) Ringdown},''
  \href{http://dx.doi.org/10.1007/s10714-018-2372-6}{{\em Gen. Rel. Grav.}
  {\bfseries 50} no.~5, (2018) 49},
\href{http://arxiv.org/abs/1801.03587}{{\ttfamily arXiv:1801.03587 [gr-qc]}}.
%%CITATION = ARXIV:1801.03587;%%.

\bibitem{Sathyaprakash:2019yqt}
B.~S. Sathyaprakash {\em et~al.}, ``{Extreme Gravity and Fundamental
  Physics},''
\href{http://arxiv.org/abs/1903.09221}{{\ttfamily arXiv:1903.09221
  [astro-ph.HE]}}.
%%CITATION = ARXIV:1903.09221;%%.

\bibitem{Hawking:1974sw}
S.~Hawking, ``{Particle Creation by Black Holes},''
  \href{http://dx.doi.org/10.1007/BF02345020}{{\em Commun. Math. Phys.}
  {\bfseries 43} (1975) 199--220}. [Erratum: Commun.Math.Phys. 46, 206 (1976)].

\bibitem{Cardoso:2019rvt}
V.~Cardoso and P.~Pani, ``{Testing the nature of dark compact objects: a status
  report},'' \href{http://dx.doi.org/10.1007/s41114-019-0020-4}{{\em Living
  Rev. Rel.} {\bfseries 22} no.~1, (2019) 4},
  \href{http://arxiv.org/abs/1904.05363}{{\ttfamily arXiv:1904.05363 [gr-qc]}}.

\bibitem{Cardoso:2014sna}
V.~Cardoso, L.~C.~B. Crispino, C.~F.~B. Macedo, H.~Okawa, and P.~Pani, ``{Light
  rings as observational evidence for event horizons: long-lived modes,
  ergoregions and nonlinear instabilities of ultracompact objects},''
  \href{http://dx.doi.org/10.1103/PhysRevD.90.044069}{{\em Phys. Rev.}
  {\bfseries D90} no.~4, (2014) 044069},
\href{http://arxiv.org/abs/1406.5510}{{\ttfamily arXiv:1406.5510 [gr-qc]}}.
%%CITATION = ARXIV:1406.5510;%%.

\bibitem{Cunha:2020azh}
P.~V. Cunha and C.~A. Herdeiro, ``{Stationary black holes and light rings},''
  \href{http://dx.doi.org/10.1103/PhysRevLett.124.181101}{{\em Phys. Rev.
  Lett.} {\bfseries 124} no.~18, (2020) 181101},
  \href{http://arxiv.org/abs/2003.06445}{{\ttfamily arXiv:2003.06445 [gr-qc]}}.

\bibitem{Hawking:1976ra}
S.~Hawking, ``{Breakdown of Predictability in Gravitational Collapse},''
  \href{http://dx.doi.org/10.1103/PhysRevD.14.2460}{{\em Phys. Rev. D}
  {\bfseries 14} (1976) 2460--2473}.

\bibitem{Lunin:2001jy}
O.~Lunin and S.~D. Mathur, ``{AdS / CFT duality and the black hole information
  paradox},'' \href{http://dx.doi.org/10.1016/S0550-3213(01)00620-4}{{\em Nucl.
  Phys. B} {\bfseries 623} (2002) 342--394},
  \href{http://arxiv.org/abs/hep-th/0109154}{{\ttfamily arXiv:hep-th/0109154}}.

\bibitem{Lunin:2002qf}
O.~Lunin and S.~D. Mathur, ``{Statistical interpretation of Bekenstein entropy
  for systems with a stretched horizon},''
  \href{http://dx.doi.org/10.1103/PhysRevLett.88.211303}{{\em Phys. Rev. Lett.}
  {\bfseries 88} (2002) 211303},
\href{http://arxiv.org/abs/hep-th/0202072}{{\ttfamily arXiv:hep-th/0202072
  [hep-th]}}.
%%CITATION = HEP-TH/0202072;%%.

\bibitem{Mathur:2005zp}
S.~D. Mathur, ``{The Fuzzball proposal for black holes: An Elementary
  review},'' \href{http://dx.doi.org/10.1002/prop.200410203}{{\em Fortsch.
  Phys.} {\bfseries 53} (2005) 793--827},
\href{http://arxiv.org/abs/hep-th/0502050}{{\ttfamily arXiv:hep-th/0502050
  [hep-th]}}.
%%CITATION = HEP-TH/0502050;%%.

\bibitem{Myers:1997qi}
R.~C. Myers, ``{Pure states don't wear black},''
  \href{http://dx.doi.org/10.1023/A:1018855611972}{{\em Gen. Rel. Grav.}
  {\bfseries 29} (1997) 1217--1222},
\href{http://arxiv.org/abs/gr-qc/9705065}{{\ttfamily arXiv:gr-qc/9705065
  [gr-qc]}}.
%%CITATION = GR-QC/9705065;%%.

\bibitem{Bena:2007kg}
I.~Bena and N.~P. Warner, ``{Black holes, black rings and their microstates},''
  \href{http://dx.doi.org/10.1007/978-3-540-79523-0_1}{{\em Lect. Notes Phys.}
  {\bfseries 755} (2008) 1--92},
\href{http://arxiv.org/abs/hep-th/0701216}{{\ttfamily arXiv:hep-th/0701216
  [hep-th]}}.
%%CITATION = HEP-TH/0701216;%%.

\bibitem{Balasubramanian:2008da}
V.~Balasubramanian, J.~de~Boer, S.~El-Showk, and I.~Messamah, ``{Black Holes as
  Effective Geometries},''
  \href{http://dx.doi.org/10.1088/0264-9381/25/21/214004}{{\em Class. Quant.
  Grav.} {\bfseries 25} (2008) 214004},
\href{http://arxiv.org/abs/0811.0263}{{\ttfamily arXiv:0811.0263 [hep-th]}}.
%%CITATION = ARXIV:0811.0263;%%.

\bibitem{Bena:2013dka}
I.~Bena and N.~P. Warner, ``{Resolving the Structure of Black Holes:
  Philosophizing with a Hammer},''
\href{http://arxiv.org/abs/1311.4538}{{\ttfamily arXiv:1311.4538 [hep-th]}}.
%%CITATION = ARXIV:1311.4538;%%.

\bibitem{Buoninfante:2019swn}
L.~Buoninfante and A.~Mazumdar, ``{Nonlocal star as a blackhole mimicker},''
  \href{http://dx.doi.org/10.1103/PhysRevD.100.024031}{{\em Phys. Rev. D}
  {\bfseries 100} no.~2, (2019) 024031},
  \href{http://arxiv.org/abs/1903.01542}{{\ttfamily arXiv:1903.01542 [gr-qc]}}.

\bibitem{Nicolini:2005vd}
P.~Nicolini, A.~Smailagic, and E.~Spallucci, ``{Noncommutative geometry
  inspired Schwarzschild black hole},''
  \href{http://dx.doi.org/10.1016/j.physletb.2005.11.004}{{\em Phys. Lett. B}
  {\bfseries 632} (2006) 547--551},
  \href{http://arxiv.org/abs/gr-qc/0510112}{{\ttfamily arXiv:gr-qc/0510112}}.

\bibitem{Koshelev:2017bxd}
A.~S. Koshelev and A.~Mazumdar, ``{Do massive compact objects without event
  horizon exist in infinite derivative gravity?},''
  \href{http://dx.doi.org/10.1103/PhysRevD.96.084069}{{\em Phys. Rev.}
  {\bfseries D96} no.~8, (2017) 084069},
\href{http://arxiv.org/abs/1707.00273}{{\ttfamily arXiv:1707.00273 [gr-qc]}}.
%%CITATION = ARXIV:1707.00273;%%.

\bibitem{Buoninfante:2018rlq}
L.~Buoninfante, A.~S. Koshelev, G.~Lambiase, J.~Marto, and A.~Mazumdar,
  ``{Conformally-flat, non-singular static metric in infinite derivative
  gravity},'' \href{http://dx.doi.org/10.1088/1475-7516/2018/06/014}{{\em JCAP}
  {\bfseries 1806} no.~06, (2018) 014},
\href{http://arxiv.org/abs/1804.08195}{{\ttfamily arXiv:1804.08195 [gr-qc]}}.
%%CITATION = ARXIV:1804.08195;%%.

\bibitem{Buoninfante:2018xif}
L.~Buoninfante, A.~S. Cornell, G.~Harmsen, A.~S. Koshelev, G.~Lambiase,
  J.~Marto, and A.~Mazumdar, ``{Towards nonsingular rotating compact object in
  ghost-free infinite derivative gravity},''
  \href{http://dx.doi.org/10.1103/PhysRevD.98.084041}{{\em Phys. Rev. D}
  {\bfseries 98} no.~8, (2018) 084041},
  \href{http://arxiv.org/abs/1807.08896}{{\ttfamily arXiv:1807.08896 [gr-qc]}}.

\bibitem{Biswas:2011ar}
T.~Biswas, E.~Gerwick, T.~Koivisto, and A.~Mazumdar, ``{Towards singularity and
  ghost free theories of gravity},''
  \href{http://dx.doi.org/10.1103/PhysRevLett.108.031101}{{\em Phys. Rev.
  Lett.} {\bfseries 108} (2012) 031101},
  \href{http://arxiv.org/abs/1110.5249}{{\ttfamily arXiv:1110.5249 [gr-qc]}}.

\bibitem{Frolov:2015bta}
V.~P. Frolov, ``{Mass-gap for black hole formation in higher derivative and
  ghost free gravity},''
  \href{http://dx.doi.org/10.1103/PhysRevLett.115.051102}{{\em Phys. Rev.
  Lett.} {\bfseries 115} no.~5, (2015) 051102},
  \href{http://arxiv.org/abs/1505.00492}{{\ttfamily arXiv:1505.00492
  [hep-th]}}.

\bibitem{Abbott:2020khf}
{\bfseries LIGO Scientific, Virgo} Collaboration, R.~Abbott {\em et~al.},
  ``{GW190814: Gravitational Waves from the Coalescence of a 23 Solar Mass
  Black Hole with a 2.6 Solar Mass Compact Object},''
  \href{http://dx.doi.org/10.3847/2041-8213/ab960f}{{\em Astrophys. J.}
  {\bfseries 896} no.~2, (2020) L44},
  \href{http://arxiv.org/abs/2006.12611}{{\ttfamily arXiv:2006.12611
  [astro-ph.HE]}}.

\bibitem{Damour:1982}
T.~{Damour}, ``{Surface Effects in Black-Hole Physics},'' in {\em Marcel
  Grossmann Meeting: General Relativity}, p.~587.
\newblock Jan., 1982.

\bibitem{Thorne:1986iy}
K.~S. Thorne, R.~Price, and D.~Macdonald, eds., {\em {BLACK HOLES: THE MEMBRANE
  PARADIGM}}.
\newblock 1986.

\bibitem{Cardoso:2016rao}
V.~Cardoso, E.~Franzin, and P.~Pani, ``{Is the gravitational-wave ringdown a
  probe of the event horizon?},''
  \href{http://dx.doi.org/10.1103/PhysRevLett.116.171101}{{\em Phys. Rev.
  Lett.} {\bfseries 116} no.~17, (2016) 171101},
\href{http://arxiv.org/abs/1602.07309}{{\ttfamily arXiv:1602.07309 [gr-qc]}}.
%%CITATION = ARXIV:1602.07309;%%.

\bibitem{Cardoso:2016oxy}
V.~Cardoso, S.~Hopper, C.~F.~B. Macedo, C.~Palenzuela, and P.~Pani,
  ``{Gravitational-wave signatures of exotic compact objects and of quantum
  corrections at the horizon scale},''
  \href{http://dx.doi.org/10.1103/PhysRevD.94.084031}{{\em Phys. Rev.}
  {\bfseries D94} no.~8, (2016) 084031},
\href{http://arxiv.org/abs/1608.08637}{{\ttfamily arXiv:1608.08637 [gr-qc]}}.
%%CITATION = ARXIV:1608.08637;%%.

\bibitem{Cardoso:2017cqb}
V.~Cardoso and P.~Pani, ``{Tests for the existence of black holes through
  gravitational wave echoes},''
  \href{http://dx.doi.org/10.1038/s41550-017-0225-y}{{\em Nat. Astron.}
  {\bfseries 1} no.~9, (2017) 586--591},
\href{http://arxiv.org/abs/1709.01525}{{\ttfamily arXiv:1709.01525 [gr-qc]}}.
%%CITATION = ARXIV:1709.01525;%%.

\bibitem{Mark:2017dnq}
Z.~Mark, A.~Zimmerman, S.~M. Du, and Y.~Chen, ``{A recipe for echoes from
  exotic compact objects},''
  \href{http://dx.doi.org/10.1103/PhysRevD.96.084002}{{\em Phys. Rev.}
  {\bfseries D96} no.~8, (2017) 084002},
\href{http://arxiv.org/abs/1706.06155}{{\ttfamily arXiv:1706.06155 [gr-qc]}}.
%%CITATION = ARXIV:1706.06155;%%.

\bibitem{Maggio:2017ivp}
E.~Maggio, P.~Pani, and V.~Ferrari, ``{Exotic Compact Objects and How to Quench
  their Ergoregion Instability},''
  \href{http://dx.doi.org/10.1103/PhysRevD.96.104047}{{\em Phys. Rev.}
  {\bfseries D96} no.~10, (2017) 104047},
\href{http://arxiv.org/abs/1703.03696}{{\ttfamily arXiv:1703.03696 [gr-qc]}}.
%%CITATION = ARXIV:1703.03696;%%.

\bibitem{Abedi:2016hgu}
J.~Abedi, H.~Dykaar, and N.~Afshordi, ``{Echoes from the Abyss: Tentative
  evidence for Planck-scale structure at black hole horizons},''
  \href{http://dx.doi.org/10.1103/PhysRevD.96.082004}{{\em Phys. Rev.}
  {\bfseries D96} no.~8, (2017) 082004},
\href{http://arxiv.org/abs/1612.00266}{{\ttfamily arXiv:1612.00266 [gr-qc]}}.
%%CITATION = ARXIV:1612.00266;%%.

\bibitem{Conklin:2017lwb}
R.~S. Conklin, B.~Holdom, and J.~Ren, ``{Gravitational wave echoes through new
  windows},'' \href{http://dx.doi.org/10.1103/PhysRevD.98.044021}{{\em Phys.
  Rev.} {\bfseries D98} no.~4, (2018) 044021},
\href{http://arxiv.org/abs/1712.06517}{{\ttfamily arXiv:1712.06517 [gr-qc]}}.
%%CITATION = ARXIV:1712.06517;%%.

\bibitem{Oshita:2018fqu}
N.~Oshita and N.~Afshordi, ``{Probing microstructure of black hole spacetimes
  with gravitational wave echoes},''
  \href{http://dx.doi.org/10.1103/PhysRevD.99.044002}{{\em Phys. Rev.}
  {\bfseries D99} no.~4, (2019) 044002},
\href{http://arxiv.org/abs/1807.10287}{{\ttfamily arXiv:1807.10287 [gr-qc]}}.
%%CITATION = ARXIV:1807.10287;%%.

\bibitem{Wang:2019rcf}
Q.~Wang, N.~Oshita, and N.~Afshordi, ``{Echoes from Quantum Black Holes},''
  \href{http://dx.doi.org/10.1103/PhysRevD.101.024031}{{\em Phys. Rev. D}
  {\bfseries 101} no.~2, (2020) 024031},
  \href{http://arxiv.org/abs/1905.00446}{{\ttfamily arXiv:1905.00446 [gr-qc]}}.

\bibitem{Cardoso:2019apo}
V.~Cardoso, V.~F. Foit, and M.~Kleban, ``{Gravitational wave echoes from black
  hole area quantization},''
  \href{http://dx.doi.org/10.1088/1475-7516/2019/08/006}{{\em JCAP} {\bfseries
  08} (2019) 006}, \href{http://arxiv.org/abs/1902.10164}{{\ttfamily
  arXiv:1902.10164 [hep-th]}}.

\bibitem{Coates:2019bun}
A.~Coates, S.~H. Völkel, and K.~D. Kokkotas, ``{Spectral Lines of Quantized,
  Spinning Black Holes and their Astrophysical Relevance},''
  \href{http://dx.doi.org/10.1103/PhysRevLett.123.171104}{{\em Phys. Rev.
  Lett.} {\bfseries 123} no.~17, (2019) 171104},
  \href{http://arxiv.org/abs/1909.01254}{{\ttfamily arXiv:1909.01254 [gr-qc]}}.

\bibitem{Buoninfante:2020tfb}
L.~Buoninfante, ``{Echoes from corpuscular black holes},''
  \href{http://arxiv.org/abs/2005.08426}{{\ttfamily arXiv:2005.08426 [gr-qc]}}.

\bibitem{Ferrari:2000sr}
V.~Ferrari and K.~D. Kokkotas, ``{Scattering of particles by neutron stars:
  Time evolutions for axial perturbations},''
  \href{http://dx.doi.org/10.1103/PhysRevD.62.107504}{{\em Phys. Rev.}
  {\bfseries D62} (2000) 107504},
\href{http://arxiv.org/abs/gr-qc/0008057}{{\ttfamily arXiv:gr-qc/0008057
  [gr-qc]}}.
%%CITATION = GR-QC/0008057;%%.

\bibitem{Pani:2018flj}
P.~Pani and V.~Ferrari, ``{On gravitational-wave echoes from neutron-star
  binary coalescences},''
  \href{http://dx.doi.org/10.1088/1361-6382/aacb8f}{{\em Class. Quant. Grav.}
  {\bfseries 35} no.~15, (2018) 15LT01},
\href{http://arxiv.org/abs/1804.01444}{{\ttfamily arXiv:1804.01444 [gr-qc]}}.
%%CITATION = ARXIV:1804.01444;%%.

\bibitem{Holdom:2016nek}
B.~Holdom and J.~Ren, ``{Not quite a black hole},''
  \href{http://dx.doi.org/10.1103/PhysRevD.95.084034}{{\em Phys. Rev.}
  {\bfseries D95} no.~8, (2017) 084034},
\href{http://arxiv.org/abs/1612.04889}{{\ttfamily arXiv:1612.04889 [gr-qc]}}.
%%CITATION = ARXIV:1612.04889;%%.

\bibitem{Buoninfante:2019teo}
L.~Buoninfante, A.~Mazumdar, and J.~Peng, ``{Nonlocality amplifies echoes},''
  \href{http://dx.doi.org/10.1103/PhysRevD.100.104059}{{\em Phys. Rev. D}
  {\bfseries 100} no.~10, (2019) 104059},
  \href{http://arxiv.org/abs/1906.03624}{{\ttfamily arXiv:1906.03624 [gr-qc]}}.

\bibitem{Delhom:2019btt}
A.~Delhom, C.~F. Macedo, G.~J. Olmo, and L.~C. Crispino, ``{Absorption by black
  hole remnants in metric-affine gravity},''
  \href{http://dx.doi.org/10.1103/PhysRevD.100.024016}{{\em Phys. Rev. D}
  {\bfseries 100} no.~2, (2019) 024016},
  \href{http://arxiv.org/abs/1906.06411}{{\ttfamily arXiv:1906.06411 [gr-qc]}}.

\bibitem{Nakano:2017fvh}
H.~Nakano, N.~Sago, H.~Tagoshi, and T.~Tanaka, ``{Black hole ringdown echoes
  and howls},'' \href{http://dx.doi.org/10.1093/ptep/ptx093}{{\em PTEP}
  {\bfseries 2017} no.~7, (2017) 071E01},
\href{http://arxiv.org/abs/1704.07175}{{\ttfamily arXiv:1704.07175 [gr-qc]}}.
%%CITATION = ARXIV:1704.07175;%%.

\bibitem{Bueno:2017hyj}
P.~Bueno, P.~A. Cano, F.~Goelen, T.~Hertog, and B.~Vercnocke, ``{Echoes of
  Kerr-like wormholes},''
  \href{http://dx.doi.org/10.1103/PhysRevD.97.024040}{{\em Phys. Rev.}
  {\bfseries D97} no.~2, (2018) 024040},
\href{http://arxiv.org/abs/1711.00391}{{\ttfamily arXiv:1711.00391 [gr-qc]}}.
%%CITATION = ARXIV:1711.00391;%%.

\bibitem{Wang:2018mlp}
Y.-T. Wang, Z.-P. Li, J.~Zhang, S.-Y. Zhou, and Y.-S. Piao, ``{Are
  gravitational wave ringdown echoes always equal-interval?},''
  \href{http://dx.doi.org/10.1140/epjc/s10052-018-5974-y}{{\em Eur. Phys. J.}
  {\bfseries C78} no.~6, (2018) 482},
\href{http://arxiv.org/abs/1802.02003}{{\ttfamily arXiv:1802.02003 [gr-qc]}}.
%%CITATION = ARXIV:1802.02003;%%.

\bibitem{Correia:2018apm}
M.~R. Correia and V.~Cardoso, ``{Characterization of echoes: A Dyson-series
  representation of individual pulses},''
  \href{http://dx.doi.org/10.1103/PhysRevD.97.084030}{{\em Phys. Rev.}
  {\bfseries D97} no.~8, (2018) 084030},
\href{http://arxiv.org/abs/1802.07735}{{\ttfamily arXiv:1802.07735 [gr-qc]}}.
%%CITATION = ARXIV:1802.07735;%%.

\bibitem{Wang:2018gin}
Q.~Wang and N.~Afshordi, ``{Black hole echology: The observer's manual},''
  \href{http://dx.doi.org/10.1103/PhysRevD.97.124044}{{\em Phys. Rev.}
  {\bfseries D97} no.~12, (2018) 124044},
\href{http://arxiv.org/abs/1803.02845}{{\ttfamily arXiv:1803.02845 [gr-qc]}}.
%%CITATION = ARXIV:1803.02845;%%.

\bibitem{Uchikata:2019frs}
N.~Uchikata, H.~Nakano, T.~Narikawa, N.~Sago, H.~Tagoshi, and T.~Tanaka,
  ``{Searching for black hole echoes from the LIGO-Virgo Catalog GWTC-1},''
  \href{http://dx.doi.org/10.1103/PhysRevD.100.062006}{{\em Phys. Rev. D}
  {\bfseries 100} no.~6, (2019) 062006},
  \href{http://arxiv.org/abs/1906.00838}{{\ttfamily arXiv:1906.00838 [gr-qc]}}.

\bibitem{Testa:2018bzd}
A.~Testa and P.~Pani, ``{Analytical template for gravitational-wave echoes:
  signal characterization and prospects of detection with current and future
  interferometers},'' \href{http://dx.doi.org/10.1103/PhysRevD.98.044018}{{\em
  Phys. Rev.} {\bfseries D98} no.~4, (2018) 044018},
\href{http://arxiv.org/abs/1806.04253}{{\ttfamily arXiv:1806.04253 [gr-qc]}}.
%%CITATION = ARXIV:1806.04253;%%.

\bibitem{Maggio:2019zyv}
E.~Maggio, A.~Testa, S.~Bhagwat, and P.~Pani, ``{Analytical model for
  gravitational-wave echoes from spinning remnants},''
  \href{http://dx.doi.org/10.1103/PhysRevD.100.064056}{{\em Phys. Rev. D}
  {\bfseries 100} no.~6, (2019) 064056},
  \href{http://arxiv.org/abs/1907.03091}{{\ttfamily arXiv:1907.03091 [gr-qc]}}.

\bibitem{Ashton:2016xff}
G.~Ashton, O.~Birnholtz, M.~Cabero, C.~Capano, T.~Dent, B.~Krishnan, G.~D.
  Meadors, A.~B. Nielsen, A.~Nitz, and J.~Westerweck, ``{Comments on: "Echoes
  from the abyss: Evidence for Planck-scale structure at black hole
  horizons"},''
\href{http://arxiv.org/abs/1612.05625}{{\ttfamily arXiv:1612.05625 [gr-qc]}}.
%%CITATION = ARXIV:1612.05625;%%.

\bibitem{Abedi:2017isz}
J.~Abedi, H.~Dykaar, and N.~Afshordi, ``{Echoes from the Abyss: The Holiday
  Edition!},''
\href{http://arxiv.org/abs/1701.03485}{{\ttfamily arXiv:1701.03485 [gr-qc]}}.
%%CITATION = ARXIV:1701.03485;%%.

\bibitem{Westerweck:2017hus}
J.~Westerweck, A.~Nielsen, O.~Fischer-Birnholtz, M.~Cabero, C.~Capano, T.~Dent,
  B.~Krishnan, G.~Meadors, and A.~H. Nitz, ``{Low significance of evidence for
  black hole echoes in gravitational wave data},''
  \href{http://dx.doi.org/10.1103/PhysRevD.97.124037}{{\em Phys. Rev.}
  {\bfseries D97} no.~12, (2018) 124037},
\href{http://arxiv.org/abs/1712.09966}{{\ttfamily arXiv:1712.09966 [gr-qc]}}.
%%CITATION = ARXIV:1712.09966;%%.

\bibitem{Abedi:2018pst}
J.~Abedi, H.~Dykaar, and N.~Afshordi, ``{Comment on: "Low significance of
  evidence for black hole echoes in gravitational wave data"},''
\href{http://arxiv.org/abs/1803.08565}{{\ttfamily arXiv:1803.08565 [gr-qc]}}.
%%CITATION = ARXIV:1803.08565;%%.

\bibitem{Conklin:2019fcs}
R.~S. Conklin and B.~Holdom, ``{Gravitational Wave "Echo" Spectra},''
\href{http://arxiv.org/abs/1905.09370}{{\ttfamily arXiv:1905.09370 [gr-qc]}}.
%%CITATION = ARXIV:1905.09370;%%.

\bibitem{Tsang:2019zra}
K.~W. Tsang, A.~Ghosh, A.~Samajdar, K.~Chatziioannou, S.~Mastrogiovanni,
  M.~Agathos, and C.~Van Den~Broeck, ``{A morphology-independent search for
  gravitational wave echoes in data from the first and second observing runs of
  Advanced LIGO and Advanced Virgo},''
  \href{http://dx.doi.org/10.1103/PhysRevD.101.064012}{{\em Phys. Rev. D}
  {\bfseries 101} no.~6, (2020) 064012},
  \href{http://arxiv.org/abs/1906.11168}{{\ttfamily arXiv:1906.11168 [gr-qc]}}.

\bibitem{Abedi:2020ujo}
J.~Abedi, N.~Afshordi, N.~Oshita, and Q.~Wang, ``{Quantum Black Holes in the
  Sky},'' \href{http://dx.doi.org/10.3390/universe6030043}{{\em Universe}
  {\bfseries 6} no.~3, (2020) 43},
  \href{http://arxiv.org/abs/2001.09553}{{\ttfamily arXiv:2001.09553 [gr-qc]}}.

\bibitem{Guo:2017jmi}
B.~Guo, S.~Hampton, and S.~D. Mathur, ``{Can we observe fuzzballs or
  firewalls?},'' \href{http://dx.doi.org/10.1007/JHEP07(2018)162}{{\em JHEP}
  {\bfseries 07} (2018) 162},
\href{http://arxiv.org/abs/1711.01617}{{\ttfamily arXiv:1711.01617 [hep-th]}}.
%%CITATION = ARXIV:1711.01617;%%.

\bibitem{Price:1986yy}
R.~Price and K.~Thorne, ``{Membrane Viewpoint on Black Holes: Properties and
  Evolution of the Stretched Horizon},''
  \href{http://dx.doi.org/10.1103/PhysRevD.33.915}{{\em Phys. Rev. D}
  {\bfseries 33} (1986) 915--941}.

\bibitem{Darmois1927}
G.~Darmois, {\em Les équations de la gravitation einsteinienne}.
\newblock Gauthier-Villars, 1927.
\newblock \url{http://eudml.org/doc/192556}.

\bibitem{Israel:1966rt}
W.~Israel, ``{Singular hypersurfaces and thin shells in general relativity},''
  \href{http://dx.doi.org/10.1007/BF02710419, 10.1007/BF02712210}{{\em Nuovo
  Cim.} {\bfseries B44S10} (1966) 1}.
[Nuovo Cim.B44,1(1966)].
%%CITATION = NUCIA,B44S10,1;%%.

\bibitem{VisserBook}
M.~Visser, {\em {Lorentzian wormholes: From Einstein to Hawking}}.
\newblock AIP, Woodbury, USA,
1996.
\newblock
%%CITATION = ISBN-9781563966538;%%.

\bibitem{Jacobson:2011dz}
T.~Jacobson, A.~Mohd, and S.~Sarkar, ``{Membrane paradigm for
  Einstein-Gauss-Bonnet gravity},''
  \href{http://dx.doi.org/10.1103/PhysRevD.95.064036}{{\em Phys. Rev. D}
  {\bfseries 95} no.~6, (2017) 064036},
  \href{http://arxiv.org/abs/1107.1260}{{\ttfamily arXiv:1107.1260 [gr-qc]}}.

\bibitem{Buchdahl:1959zz}
H.~A. Buchdahl, ``{General Relativistic Fluid Spheres},''
\href{http://dx.doi.org/10.1103/PhysRev.116.1027}{{\em Phys. Rev.} {\bfseries
  116} (1959) 1027}.
%%CITATION = PHRVA,116,1027;%%.

\bibitem{Oshita:2019sat}
N.~Oshita, Q.~Wang, and N.~Afshordi, ``{On Reflectivity of Quantum Black Hole
  Horizons},'' \href{http://dx.doi.org/10.1088/1475-7516/2020/04/016}{{\em
  JCAP} {\bfseries 04} (2020) 016},
  \href{http://arxiv.org/abs/1905.00464}{{\ttfamily arXiv:1905.00464
  [hep-th]}}.

\bibitem{Regge:1957td}
T.~Regge and J.~A. Wheeler, ``{Stability of a Schwarzschild singularity},''
  \href{http://dx.doi.org/10.1103/PhysRev.108.1063}{{\em Phys. Rev.} {\bfseries
  108} (1957) 1063--1069}.

\bibitem{Zerilli:1970se}
F.~J. Zerilli, ``{Effective potential for even parity Regge-Wheeler
  gravitational perturbation equations},''
  \href{http://dx.doi.org/10.1103/PhysRevLett.24.737}{{\em Phys. Rev. Lett.}
  {\bfseries 24} (1970) 737--738}.

\bibitem{Zerilli:1971wd}
F.~Zerilli, ``{Gravitational field of a particle falling in a schwarzschild
  geometry analyzed in tensor harmonics},''
\href{http://dx.doi.org/10.1103/PhysRevD.2.2141}{{\em Phys. Rev.} {\bfseries
  D2} (1970) 2141--2160}.
%%CITATION = PHRVA,D2,2141;%%.

\bibitem{Maggio:2018ivz}
E.~Maggio, V.~Cardoso, S.~R. Dolan, and P.~Pani, ``{Ergoregion instability of
  exotic compact objects: electromagnetic and gravitational perturbations and
  the role of absorption},''
  \href{http://dx.doi.org/10.1103/PhysRevD.99.064007}{{\em Phys. Rev.}
  {\bfseries D99} no.~6, (2019) 064007},
\href{http://arxiv.org/abs/1807.08840}{{\ttfamily arXiv:1807.08840 [gr-qc]}}.
%%CITATION = ARXIV:1807.08840;%%.

\bibitem{Brito:2015oca}
R.~Brito, V.~Cardoso, and P.~Pani, ``{Superradiance},''
  \href{http://dx.doi.org/10.1007/978-3-319-19000-6}{{\em Lect. Notes Phys.}
  {\bfseries 906} (2015) pp.1--237},
\href{http://arxiv.org/abs/1501.06570}{{\ttfamily arXiv:1501.06570 [gr-qc]}}.
%%CITATION = ARXIV:1501.06570;%%.

\bibitem{Visser:2003ge}
M.~Visser and D.~L. Wiltshire, ``{Stable gravastars: An Alternative to black
  holes?},'' \href{http://dx.doi.org/10.1088/0264-9381/21/4/027}{{\em Class.
  Quant. Grav.} {\bfseries 21} (2004) 1135--1152},
\href{http://arxiv.org/abs/gr-qc/0310107}{{\ttfamily arXiv:gr-qc/0310107
  [gr-qc]}}.
%%CITATION = GR-QC/0310107;%%.

\bibitem{Pani:2010em}
P.~Pani, E.~Berti, V.~Cardoso, Y.~Chen, and R.~Norte, ``{Gravitational-wave
  signatures of the absence of an event horizon. II. Extreme mass ratio
  inspirals in the spacetime of a thin-shell gravastar},''
  \href{http://dx.doi.org/10.1103/PhysRevD.81.084011}{{\em Phys. Rev.}
  {\bfseries D81} (2010) 084011},
\href{http://arxiv.org/abs/1001.3031}{{\ttfamily arXiv:1001.3031 [gr-qc]}}.
%%CITATION = ARXIV:1001.3031;%%.

\bibitem{Pani:2013pma}
P.~Pani, ``{Advanced Methods in Black-Hole Perturbation Theory},''
  \href{http://dx.doi.org/10.1142/S0217751X13400186}{{\em Int. J. Mod. Phys.}
  {\bfseries A28} (2013) 1340018},
\href{http://arxiv.org/abs/1305.6759}{{\ttfamily arXiv:1305.6759 [gr-qc]}}.
%%CITATION = ARXIV:1305.6759;%%.

\bibitem{Leaver:1985ax}
E.~Leaver, ``{An Analytic representation for the quasi normal modes of Kerr
  black holes},'' \href{http://dx.doi.org/10.1098/rspa.1985.0119}{{\em Proc.
  Roy. Soc. Lond. A} {\bfseries A402} (1985) 285--298}.

\bibitem{Leins:1993zz}
M.~Leins, H.~Nollert, and M.~Soffel, ``{Nonradial oscillations of neutron
  stars: A New branch of strongly damped normal modes},''
  \href{http://dx.doi.org/10.1103/PhysRevD.48.3467}{{\em Phys. Rev. D}
  {\bfseries 48} (1993) 3467--3472}.

\bibitem{Benhar:1998au}
O.~Benhar, E.~Berti, and V.~Ferrari, ``{The Imprint of the equation of state on
  the axial w modes of oscillating neutron stars},''
  \href{http://dx.doi.org/10.1046/j.1365-8711.1999.02983.x}{{\em ICTP Lect.
  Notes Ser.} {\bfseries 3} (2001) 35--46},
  \href{http://arxiv.org/abs/gr-qc/9901037}{{\ttfamily arXiv:gr-qc/9901037}}.

\bibitem{Pani:2009ss}
P.~Pani, E.~Berti, V.~Cardoso, Y.~Chen, and R.~Norte, ``{Gravitational wave
  signatures of the absence of an event horizon. I. Nonradial oscillations of a
  thin-shell gravastar},''
  \href{http://dx.doi.org/10.1103/PhysRevD.80.124047}{{\em Phys. Rev.}
  {\bfseries D80} (2009) 124047},
\href{http://arxiv.org/abs/0909.0287}{{\ttfamily arXiv:0909.0287 [gr-qc]}}.
%%CITATION = ARXIV:0909.0287;%%.

\bibitem{Chandrasekhar:1975zza}
S.~Chandrasekhar and S.~L. Detweiler, ``{The quasi-normal modes of the
  Schwarzschild black hole},''
  \href{http://dx.doi.org/10.1098/rspa.1975.0112}{{\em Proc. Roy. Soc. Lond. A}
  {\bfseries A344} (1975) 441--452}.

\bibitem{Giesler:2019uxc}
M.~Giesler, M.~Isi, M.~A. Scheel, and S.~Teukolsky, ``{Black Hole Ringdown: The
  Importance of Overtones},''
  \href{http://dx.doi.org/10.1103/PhysRevX.9.041060}{{\em Phys. Rev. X}
  {\bfseries 9} no.~4, (2019) 041060},
  \href{http://arxiv.org/abs/1903.08284}{{\ttfamily arXiv:1903.08284 [gr-qc]}}.

\bibitem{Isi:2019aib}
M.~Isi, M.~Giesler, W.~M. Farr, M.~A. Scheel, and S.~A. Teukolsky, ``{Testing
  the no-hair theorem with GW150914},''
  \href{http://dx.doi.org/10.1103/PhysRevLett.123.111102}{{\em Phys. Rev.
  Lett.} {\bfseries 123} no.~11, (2019) 111102},
  \href{http://arxiv.org/abs/1905.00869}{{\ttfamily arXiv:1905.00869 [gr-qc]}}.

\bibitem{Berti:2007zu}
E.~Berti, J.~Cardoso, V.~Cardoso, and M.~Cavaglia, ``{Matched-filtering and
  parameter estimation of ringdown waveforms},''
  \href{http://dx.doi.org/10.1103/PhysRevD.76.104044}{{\em Phys. Rev.}
  {\bfseries D76} (2007) 104044},
\href{http://arxiv.org/abs/0707.1202}{{\ttfamily arXiv:0707.1202 [gr-qc]}}.
%%CITATION = ARXIV:0707.1202;%%.

\bibitem{Baibhav:2017jhs}
V.~Baibhav, E.~Berti, V.~Cardoso, and G.~Khanna, ``{Black Hole Spectroscopy:
  Systematic Errors and Ringdown Energy Estimates},''
  \href{http://dx.doi.org/10.1103/PhysRevD.97.044048}{{\em Phys. Rev.}
  {\bfseries D97} no.~4, (2018) 044048},
\href{http://arxiv.org/abs/1710.02156}{{\ttfamily arXiv:1710.02156 [gr-qc]}}.
%%CITATION = ARXIV:1710.02156;%%.

\bibitem{Bhagwat:2019dtm}
S.~Bhagwat, X.~J. Forteza, P.~Pani, and V.~Ferrari, ``{Ringdown overtones,
  black hole spectroscopy, and no-hair theorem tests},''
  \href{http://dx.doi.org/10.1103/PhysRevD.101.044033}{{\em Phys. Rev. D}
  {\bfseries 101} no.~4, (2020) 044033},
  \href{http://arxiv.org/abs/1910.08708}{{\ttfamily arXiv:1910.08708 [gr-qc]}}.

\bibitem{Ota:2019bzl}
I.~Ota and C.~Chirenti, ``{Overtones or higher harmonics? Prospects for testing
  the no-hair theorem with gravitational wave detections},''
  \href{http://dx.doi.org/10.1103/PhysRevD.101.104005}{{\em Phys. Rev. D}
  {\bfseries 101} no.~10, (2020) 104005},
  \href{http://arxiv.org/abs/1911.00440}{{\ttfamily arXiv:1911.00440 [gr-qc]}}.

\bibitem{Forteza:2020hbw}
X.~J. Forteza, S.~Bhagwat, P.~Pani, and V.~Ferrari, ``{On the spectroscopy of
  binary black hole ringdown using overtones and angular modes},''
  \href{http://arxiv.org/abs/2005.03260}{{\ttfamily arXiv:2005.03260 [gr-qc]}}.

\bibitem{Nollert:1996rf}
H.-P. Nollert, ``{About the significance of quasinormal modes of black
  holes},'' \href{http://dx.doi.org/10.1103/PhysRevD.53.4397}{{\em Phys. Rev.
  D} {\bfseries 53} (1996) 4397--4402},
  \href{http://arxiv.org/abs/gr-qc/9602032}{{\ttfamily arXiv:gr-qc/9602032}}.

\bibitem{Leung:1999iq}
P.~Leung, Y.~Liu, W.~Suen, C.~Tam, and K.~Young, ``{Perturbative approach to
  the quasinormal modes of dirty black holes},''
  \href{http://dx.doi.org/10.1103/PhysRevD.59.044034}{{\em Phys. Rev. D}
  {\bfseries 59} (1999) 044034},
  \href{http://arxiv.org/abs/gr-qc/9903032}{{\ttfamily arXiv:gr-qc/9903032}}.

\bibitem{Barausse:2014tra}
E.~Barausse, V.~Cardoso, and P.~Pani, ``{Can environmental effects spoil
  precision gravitational-wave astrophysics?},''
  \href{http://dx.doi.org/10.1103/PhysRevD.89.104059}{{\em Phys. Rev.}
  {\bfseries D89} no.~10, (2014) 104059},
\href{http://arxiv.org/abs/1404.7149}{{\ttfamily arXiv:1404.7149 [gr-qc]}}.
%%CITATION = ARXIV:1404.7149;%%.

\bibitem{Jaramillo:2020tuu}
J.~L. Jaramillo, R.~Panosso~Macedo, and L.~Al~Sheikh, ``{Pseudospectrum and
  black hole quasi-normal mode (in)stability},''
  \href{http://arxiv.org/abs/2004.06434}{{\ttfamily arXiv:2004.06434 [gr-qc]}}.

\bibitem{Hild:2010id}
S.~Hild {\em et~al.}, ``{Sensitivity Studies for Third-Generation Gravitational
  Wave Observatories},''
  \href{http://dx.doi.org/10.1088/0264-9381/28/9/094013}{{\em Class. Quant.
  Grav.} {\bfseries 28} (2011) 094013},
\href{http://arxiv.org/abs/1012.0908}{{\ttfamily arXiv:1012.0908 [gr-qc]}}.
%%CITATION = ARXIV:1012.0908;%%.

\bibitem{Audley:2017drz}
H.~{Audley}, S.~{Babak}, J.~{Baker}, E.~{Barausse}, P.~{Bender}, E.~{Berti},
  P.~{Binetruy}, M.~{Born}, D.~{Bortoluzzi}, J.~{Camp}, C.~{Caprini},
  V.~{Cardoso}, M.~{Colpi}, J.~{Conklin}, N.~{Cornish}, C.~{Cutler}, {\em
  et~al.}, ``{Laser Interferometer Space Antenna},'' {\em ArXiv e-prints}
  (Feb., 2017) , \href{http://arxiv.org/abs/1702.00786}{{\ttfamily
  arXiv:1702.00786 [astro-ph.IM]}}.

\bibitem{Baibhav:2019rsa}
V.~Baibhav {\em et~al.}, ``{Probing the Nature of Black Holes: Deep in the mHz
  Gravitational-Wave Sky},'' \href{http://arxiv.org/abs/1908.11390}{{\ttfamily
  arXiv:1908.11390 [astro-ph.HE]}}.

\bibitem{Damour:2007ap}
T.~Damour and S.~N. Solodukhin, ``{Wormholes as black hole foils},''
  \href{http://dx.doi.org/10.1103/PhysRevD.76.024016}{{\em Phys. Rev.}
  {\bfseries D76} (2007) 024016},
\href{http://arxiv.org/abs/0704.2667}{{\ttfamily arXiv:0704.2667 [gr-qc]}}.
%%CITATION = ARXIV:0704.2667;%%.

\bibitem{Mazur:2004fk}
P.~O. Mazur and E.~Mottola, ``{Gravitational vacuum condensate stars},''
  \href{http://dx.doi.org/10.1073/pnas.0402717101}{{\em Proc. Nat. Acad. Sci.}
  {\bfseries 101} (2004) 9545--9550},
\href{http://arxiv.org/abs/gr-qc/0407075}{{\ttfamily arXiv:gr-qc/0407075
  [gr-qc]}}.
%%CITATION = GR-QC/0407075;%%.

\bibitem{Chirenti:2007mk}
C.~B. M.~H. Chirenti and L.~Rezzolla, ``{How to tell a gravastar from a black
  hole},'' \href{http://dx.doi.org/10.1088/0264-9381/24/16/013}{{\em Class.
  Quant. Grav.} {\bfseries 24} (2007) 4191--4206},
\href{http://arxiv.org/abs/0706.1513}{{\ttfamily arXiv:0706.1513 [gr-qc]}}.
%%CITATION = ARXIV:0706.1513;%%.

\bibitem{Gimon:2007ur}
E.~G. Gimon and P.~Horava, ``{Astrophysical violations of the Kerr bound as a
  possible signature of string theory},''
  \href{http://dx.doi.org/10.1016/j.physletb.2009.01.026}{{\em Phys. Lett.}
  {\bfseries B672} (2009) 299--302},
\href{http://arxiv.org/abs/0706.2873}{{\ttfamily arXiv:0706.2873 [hep-th]}}.
%%CITATION = ARXIV:0706.2873;%%.

\bibitem{Lousto:2002em}
C.~O. Lousto and B.~F. Whiting, ``{Reconstruction of black hole metric
  perturbations from Weyl curvature},''
  \href{http://dx.doi.org/10.1103/PhysRevD.66.024026}{{\em Phys. Rev. D}
  {\bfseries 66} (2002) 024026},
  \href{http://arxiv.org/abs/gr-qc/0203061}{{\ttfamily arXiv:gr-qc/0203061}}.

\bibitem{Whiting:2005hr}
B.~Whiting and L.~Price, ``{Metric reconstruction from Weyl scalars},''
  \href{http://dx.doi.org/10.1088/0264-9381/22/15/003}{{\em Class. Quant.
  Grav.} {\bfseries 22} (2005) S589--S604}.

\bibitem{Raposo:2018xkf}
G.~Raposo, P.~Pani, and R.~Emparan, ``{Exotic compact objects with soft
  hair},'' \href{http://dx.doi.org/10.1103/PhysRevD.99.104050}{{\em Phys. Rev.}
  {\bfseries D99} no.~10, (2019) 104050},
\href{http://arxiv.org/abs/1812.07615}{{\ttfamily arXiv:1812.07615 [gr-qc]}}.
%%CITATION = ARXIV:1812.07615;%%.

\bibitem{Cardoso:2007az}
V.~Cardoso, P.~Pani, M.~Cadoni, and M.~Cavaglia, ``{Ergoregion instability of
  ultracompact astrophysical objects},''
  \href{http://dx.doi.org/10.1103/PhysRevD.77.124044}{{\em Phys. Rev.}
  {\bfseries D77} (2008) 124044},
\href{http://arxiv.org/abs/0709.0532}{{\ttfamily arXiv:0709.0532 [gr-qc]}}.
%%CITATION = ARXIV:0709.0532;%%.

\bibitem{Cardoso:2008kj}
V.~Cardoso, P.~Pani, M.~Cadoni, and M.~Cavaglia, ``{Instability of
  hyper-compact Kerr-like objects},''
  \href{http://dx.doi.org/10.1088/0264-9381/25/19/195010}{{\em Class. Quant.
  Grav.} {\bfseries 25} (2008) 195010},
\href{http://arxiv.org/abs/0808.1615}{{\ttfamily arXiv:0808.1615 [gr-qc]}}.
%%CITATION = ARXIV:0808.1615;%%.

\bibitem{Pani:2010jz}
P.~Pani, E.~Barausse, E.~Berti, and V.~Cardoso, ``{Gravitational instabilities
  of superspinars},'' \href{http://dx.doi.org/10.1103/PhysRevD.82.044009}{{\em
  Phys. Rev.} {\bfseries D82} (2010) 044009},
\href{http://arxiv.org/abs/1006.1863}{{\ttfamily arXiv:1006.1863 [gr-qc]}}.
%%CITATION = ARXIV:1006.1863;%%.

\bibitem{Kojima:1992ie}
Y.~Kojima, ``{Equations governing the nonradial oscillations of a slowly
  rotating relativistic star},''
  \href{http://dx.doi.org/10.1103/PhysRevD.46.4289}{{\em Phys. Rev. D}
  {\bfseries 46} (1992) 4289--4303}.

\bibitem{Uchikata:2016qku}
N.~Uchikata, S.~Yoshida, and P.~Pani, ``{Tidal deformability and I-Love-Q
  relations for gravastars with polytropic thin shells},''
  \href{http://dx.doi.org/10.1103/PhysRevD.94.064015}{{\em Phys. Rev.}
  {\bfseries D94} no.~6, (2016) 064015},
\href{http://arxiv.org/abs/1607.03593}{{\ttfamily arXiv:1607.03593 [gr-qc]}}.
%%CITATION = ARXIV:1607.03593;%%.

\end{thebibliography}\endgroup
%----------------------------------------------------------------------------------------------------

\end{document}